\documentclass[reprint,twocolumn,superscriptaddress]{revtex4-1}
\usepackage{color}
\usepackage[colorlinks=true,linkcolor=blue,citecolor=blue,urlcolor = blue]{hyperref}
\usepackage{mathtools} 
\usepackage{amsmath}
\usepackage{graphicx}
\usepackage{lineno}
\usepackage{braket} 
\usepackage{dcolumn} 
\usepackage{siunitx}
\usepackage{booktabs}
\usepackage{lipsum}
\usepackage{amsfonts}

\usepackage{hyperref}
\begin{document}

\begin{abstract}
We study three-atom inelastic scattering in ultracold $\prescript{39}{}{\mathrm{K}}$ near a Feshbach resonance of intermediate coupling strength. The non-universal character of such resonance leads to an abnormally large Efimov absolute length scale and a relatively small effective range $r_e$, allowing the features of the $\prescript{39}{}{\mathrm{K}}$ Efimov spectrum to be better isolated from the short-range physics. Meticulous characterization of and correction for finite temperature effects ensure high accuracy on the measurements of these features at large-magnitude scattering lengths. For a single Feshbach resonance, we unambiguously locate four distinct features in the Efimov structure. Three of these features form ratios that obey the Efimov universal scaling to within 10\%, while the fourth feature, occurring at a value of scattering length closest to $r_e$, instead deviates from the universal value.
\end{abstract}

\title{Observation of Efimov Universality across a Non-Universal Feshbach Resonance in $\prescript{39}{}{\mathrm{K}}$}

\author{Xin Xie}
\affiliation{JILA, National Institute of Standards and Technology and the University of Colorado, and Department of Physics, Boulder, Colorado 80309, USA}
\author{Michael J. Van de Graaff}
\affiliation{JILA, National Institute of Standards and Technology and the University of Colorado, and Department of Physics, Boulder, Colorado 80309, USA}
\author{Roman Chapurin}
\affiliation{JILA, National Institute of Standards and Technology and the University of Colorado, and Department of Physics, Boulder, Colorado 80309, USA}
\author{Matthew D. Frye}
\affiliation{Joint Quantum Centre (JQC) Durham-Newcastle, Department of Chemistry, Durham University, South Road, Durham DH1 3LE, United Kingdom}
\author{Jeremy M. Hutson}
\affiliation{Joint Quantum Centre (JQC) Durham-Newcastle, Department of Chemistry, Durham University, South Road, Durham DH1 3LE, United Kingdom}
\author{Jos\'e P. D'Incao}
\affiliation{JILA, National Institute of Standards and Technology and the University of Colorado, and Department of Physics, Boulder, Colorado 80309, USA}
\author{Paul S. Julienne}
\affiliation{Joint Quantum Institute, National Institute of Standards and Technology and the University of Maryland, College Park, MD 20742}
\author{Jun Ye}
\affiliation{JILA, National Institute of Standards and Technology and the University of Colorado, and Department of Physics, Boulder, Colorado 80309, USA}
\author{Eric A. Cornell}
\affiliation{JILA, National Institute of Standards and Technology and the University of Colorado, and Department of Physics, Boulder, Colorado 80309, USA}

\date{\today}
\maketitle

Physics has always been about explaining a lot with a little.  From single-particle harmonic oscillators to critical exponents in many-body physics, we look for parsimonious descriptions and simple patterns that are universal over a huge range of energy and length scales, independent of the details of the system. The motion of three bodies, especially when of comparable masses, is famously unamenable to the application of such universal ideas. A dramatic exception occurs when the system is characterized by pairwise interactions which are near-resonant, i.e. with the \textit{s}-wave scattering length $a$ large compared to the range of the pairwise physical potential. This is the realm of Efimov physics \cite{efimov1970plb, braaten2006pr, greene2017rmp, dincao2018jpb}.

\begin{figure*}
\includegraphics[width=16cm,keepaspectratio]{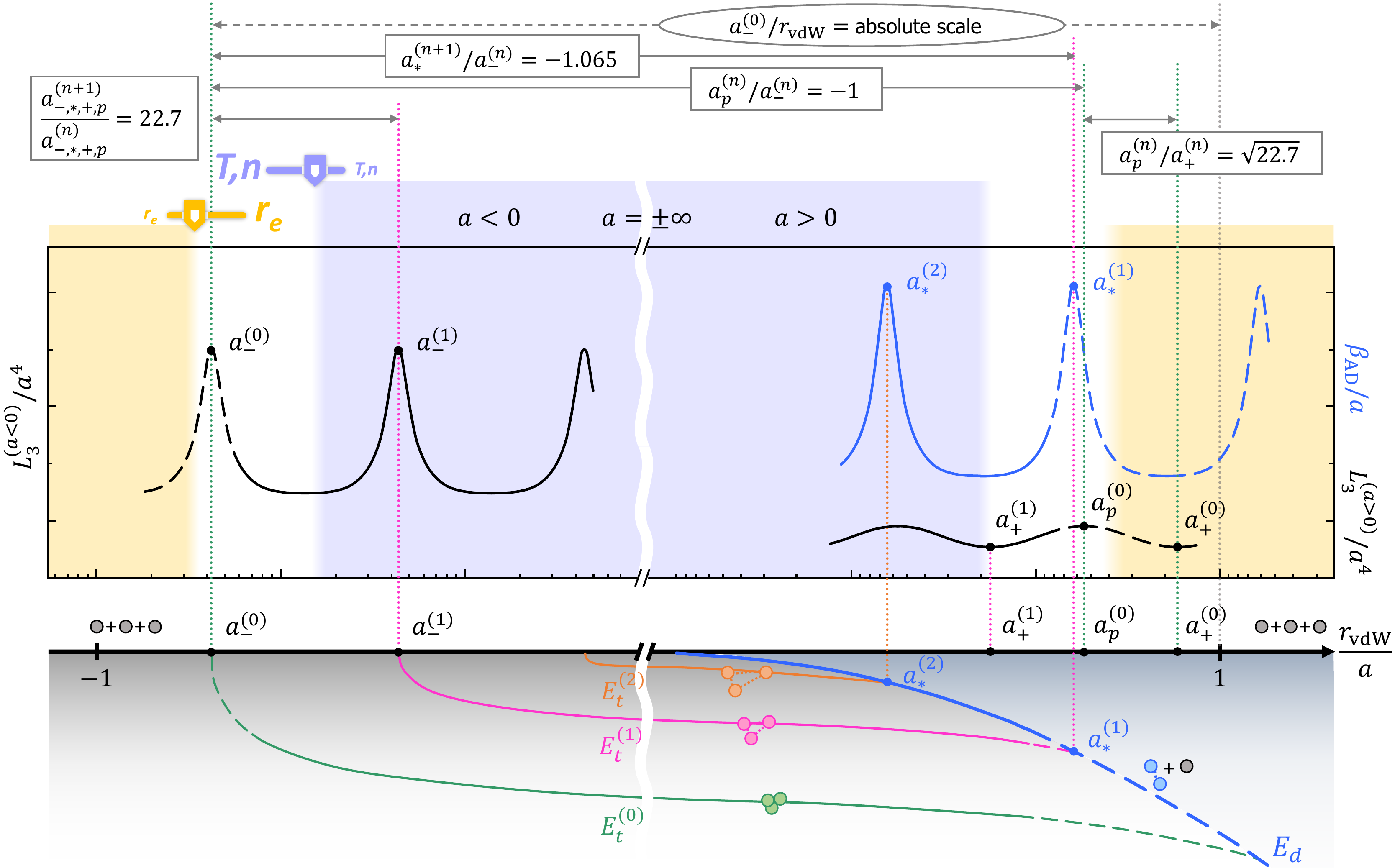} 
\centering
\caption{Efimov universality. The bottom panel shows trimer ($E_{t}^{(n)}$) and dimer ($E_{d}$) energies as a function of $a^{-1}$, with energy levels distorted to make clearer the resonance locations. The $n$th trimer is resonant with three free atoms at $a = a_{-}^{(n)}$, and with the dimer at $a = a_{*}^{(n)}$. The middle panel depicts the rate coefficients for inelastic collisions. The three-body recombination coefficient $L_3$ peaks at each value of $a_{-}^{(n)}$, while the atom-dimer relaxation coefficient $\beta_\mathrm{AD}$ peaks at each value of $a_{*}^{(n)}$. The entire structure is log-periodic with period 22.7. For $a>0$, $L_3/a^4$ shows quantum interference, with each local maximum $a_{p}^{(n)}$ spaced from the corresponding local minimum $a_{+}^{(n)}$ by $\sqrt{22.7}$. Each value of $a_{p}^{(n)}$ is related to the corresponding $a_{-}^{(n)}$ by a factor of exactly $-1$, but is offset from the nearest value of $a_{*}$ by 6.5\%. Going from relative to absolute values of $a$ requires a single absolute scale indicated by the oval circle. All these ratios are for the zero-range limit; the dashed curves suggest the possibility of perturbations as $\left| a \right|$ enters the region, indicated by yellow shading, where it is not large compared to the effective range $r_e$. The range of the yellow zone may be ``adjusted'' by choosing atomic species with different $r_e$. The purple shading represents regions of large $\left| a \right|$ that are prone to systematic effects such as those caused by finite temperature and density.
}
\label{fig:cartoon} 
\end{figure*}

Three identical \cite{heteronuclear} bosons have been shown theoretically to support an infinite sequence of three-body bound states, the Efimov states, whose spectrum as a function of $a$ displays discrete scaling invariance \cite{efimov1970plb, braaten2006pr}. The lovely recursive pattern of energy levels and associated log-periodic structure of three-body observables -- inelastic collision rates -- are depicted schematically in Fig.~\ref{fig:cartoon}. Features in the spectrum associated with successive generations of Efimov states form ratios of 22.7, while features on opposite sides of the resonance occur at a ratio of either $-1.000$ or $-1.065$ \cite{braaten2006pr, gogolin2008prl, helfrich2010pra, dincao2018jpb}. In the ideal case, these universal ratios identify the location of every feature up to a single absolute length scale. Experimental observation of this universal structure of Efimovian scaling has been a challenge, with relevant earlier measurements reviewed in the discussion near the end of this Letter. As for the absolute length scale, it was originally relegated to the status of species-dependent non-universal details \cite{dincao2009jpb}. Later it was empirically \cite{berninger2011prl, wild2012prl} and then theoretically \cite{wang2012prl, schmidt2012epjb, naidon2014prl, langmack2018pra} shown that in ultracold atom systems the absolute scale, as specified by the actual value of $a_{-}^{(0)}$, is often within 15 percent of $-9.7 \thinspace r_\mathrm{vdW}$, where $r_\mathrm{vdW}$ is the van der Waals length \cite{chin2010rmp}. This ``van der Waals universality'' is by itself remarkable, but does not speak to the original notion of Efimov universality, which is about the \textit{relative} location of multiple three-body features near a given two-body resonance.

%% figure 1

What makes observations of the originally conceived Efimov universality \cite{efimov1970plb, braaten2006pr} so difficult is a sandwiching effect.  Universality assumes that $a$ is tuned by an idealized zero-range two-body resonance, whereas any realistic scattering process is parameterized by an effective range $r_e$ \cite{scatt}, the lowest-order correction to the zero-range  approximation. Efimov features that appear at those values of $a$ (yellow shaded zone in Fig.~\ref{fig:cartoon}) that are so small as to be not well-separated from $r_e$ can be perturbed by short-range details. On the other hand, if $a$ is too large (purple shaded zone in Fig.~\ref{fig:cartoon}),  finite-temperature and finite-density effects in a bulk gas will obscure the features. Counterintuitively, van der Waals universality is the enemy of Efimov universality. The ``universal'' value of $a_{-}^{(0)}$ in principle arises from strong Feshbach resonances and is relatively small in magnitude \cite{chin2011arxiv, wang2012prl, schmidt2012epjb, langmack2018pra}. A strong resonance moreover gives rise to a large value of $r_e \approx 2.8 \thinspace r_\mathrm{vdW}$ \cite{gao1998pra, gao2011pra}, causing experimentally accessible features in the Efimov spectrum to be perilously close to the yellow zone.

In this Letter, we instead work with an intermediate-strength \cite{chapurin2019prl} Feshbach resonance in $\prescript{39}{}{\mathrm{K}}$. Too weak to obey van der Waals universality, our resonance gives rise to a correspondingly smaller $r_e$, and a larger magnitude $a_{-}^{(0)}$. The Efimov features hence tend to occur at higher values of $\left| a \right|/r_e$, away from the yellow zone. We have carefully characterized four distinct Efimov features associated with a single Feshbach resonance, an unprecedented achievement. For each measured feature, we study the temperature dependence of its location in order to extract a $T \to 0$ value, thus minimizing the hazard represented by the purple zone. These four locations
yield three independent ratios which give a measure of redundancy (see Fig.~\ref{fig:cartoon}). We identify three of these features which are arranged in Efimov universal ratios and one (the one at the lowest value of $\left| a \right|$) which is distinctly nonuniversal. Further confidence in our experimental observables comes from excellent agreement with our theoretical analysis, performed using a complete two-body coupled-channel model and a realistic three-body model built upon hyperspherical adiabatic representation, which incorporates the proper hyperfine structure and a variable number of singlet and triplet two-body bound states \cite{wang2011pra, chapurin2019prl, SuppMat, dincaoinprep}.

\begin{figure}
\includegraphics[width=8.5cm,keepaspectratio]{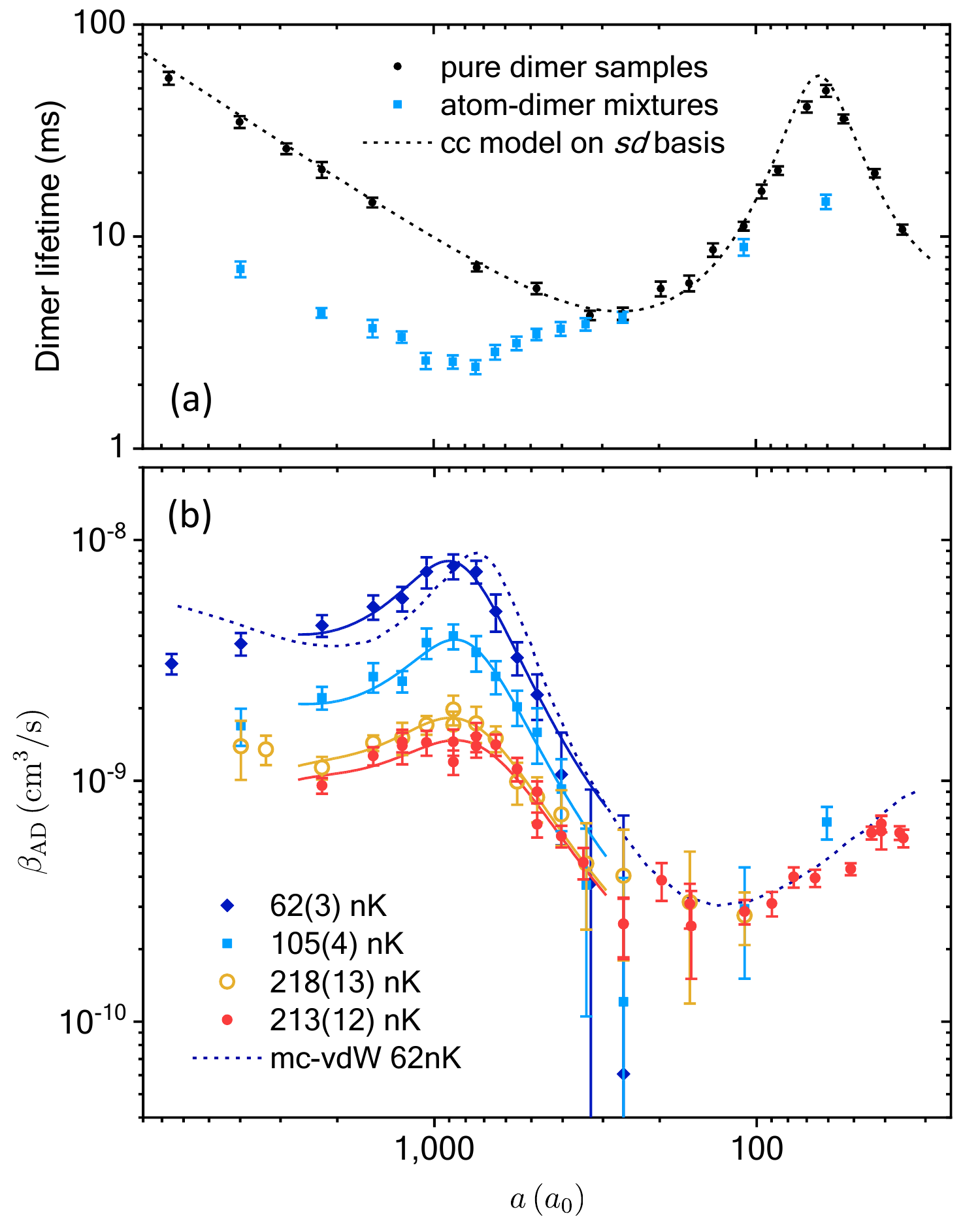} 
\centering
\caption{Temperature dependence of atom-dimer relaxation coefficient $\beta_{\mathrm{AD}}$ as a function of $a$. (a) Lifetimes of dimers with and without atoms being present. The black circles represent the intrinsic lifetimes of the dimers measured on dimer samples with $\langle n_\mathrm{D} \rangle = 2 \times 10^9 \, \mathrm{cm^{-3}}$, $T = 70 \, \mathrm{nK}$. Error bars are extracted from the fitting routine and include only the statistical noise on dimer number. The dashed line represents a coupled-channel model that includes spin--spin dipolar interaction plus second-order spin--orbit coupling \cite{SuppMat}. (b) $\beta_{\mathrm{AD}}$ measured at various temperatures. Atomic densities differ by a factor of 3 between the two highest temperature data sets. Error bars stand for $1 \sigma$ propagated uncertainty involving the statistical error of atom density as well as the uncertainty of dimer lifetimes. Solid lines are fitting curves with a finite-temperature model \cite{helfrich2009epl}. The navy dashed line is an independent prediction of our three-body multi-channel (mc-vdW) model with no adjustable parameters at $62 \thinspace \mathrm{nK}$, obtained using 4 (3) \textit{s}-wave singlet (triplet) two-body bound states \cite{SuppMat}.
}
\label{fig:atomdimer} 
\end{figure}

%% AD loss
The Efimov feature for $a<0$ (the tri-atomic resonance) was studied in our previous work \cite{chapurin2019prl}. Here, for $a>0$, we first discuss the atom-dimer scattering resonance, which manifests as enhanced atom-dimer inelastic decay rate, $\beta_\mathrm{AD}$. After the evaporative cooling in a pancake-shaped crossed dipole trap, we end up with a spin-polarized cloud at various temperatures. An admixture of atoms and dimers is generated by magnetoassociation of the dimers and followed by a step to precisely control the atom density. We then hold the samples at different magnetic fields and track the populations of the dimers as they either break apart on their own or react with free atoms. The dimers in non-ground Zeeman sublevels are known to dissociate spontaneously and spin-flip into \textit{d}-wave exit channels \cite{thompson2005prl, kohler2005prl}. This decay process contributes to a background loss of dimers as shown in Fig.~\ref{fig:atomdimer}(a). The pure dimer lifetime shows a peak around $a = 65 \thinspace a_0$. We calculate this lifetime using coupled-channel methods \cite{molscat:2019, mbf-github:2020, Frye:quasibound:2020} and find that the position and height of the peak are very sensitive to interference between different \textit{d}-wave decay paths. We determine the strength of the second-order spin--orbit coupling, which was previously neglected \cite{Falke:2008} but significantly influences the balance between paths here \cite{SuppMat}. The dashed line in Fig.~\ref{fig:atomdimer}(a) shows the resulting theoretical curve. On top of this two-body inelastic process, we observe that the dimers become shorter-lived due to their reaction with atoms. By subtracting out the dimer one-body decay rate from the dimer total decay rate \cite{SuppMat}, we extract the atom-dimer relaxation coefficient $\beta_{\mathrm{AD}}$ at different temperatures as plotted in Fig.~\ref{fig:atomdimer}(b). A resonant peak is pronounced at all temperatures. The highest temperature data were collected with multiple atom densities in order to verify the negligible role of four-body processes.

%% figure 2

To quantitatively study the resonant behavior of $\beta_{\mathrm{AD}}$, we fit the data [Fig.~\ref{fig:atomdimer}(b)] with a zero-range effective field theory \cite{helfrich2009epl} that provides a convenient parametrization of atom-dimer scattering at finite energy. There are two free parameters in this model, $a_{*}$ and $\eta_{*}$ \cite{SuppMat}. $a_{*}$ denotes the position of the resonance where an Efimov state merges into the atom-dimer scattering threshold; $\eta_{*}$ is the inelasticity parameter that characterizes the probability of decay into an energetic atom and deep dimer. We include an additional parameter in the fitting function, the global magnitude $A_*$, which serves as a diagnostic indicator of the overall consistency between experiment and theory. The temperature of the sample, which is an input parameter to this model, is measured with absorption images on atoms after a long time of flight. As depicted by the set of solid lines, this finite temperature model captures the shape of the atom-dimer resonance peak across the whole temperature range accessed in our experiment.

%% figure 3
\begin{figure}
\includegraphics[width=8.65cm,keepaspectratio]{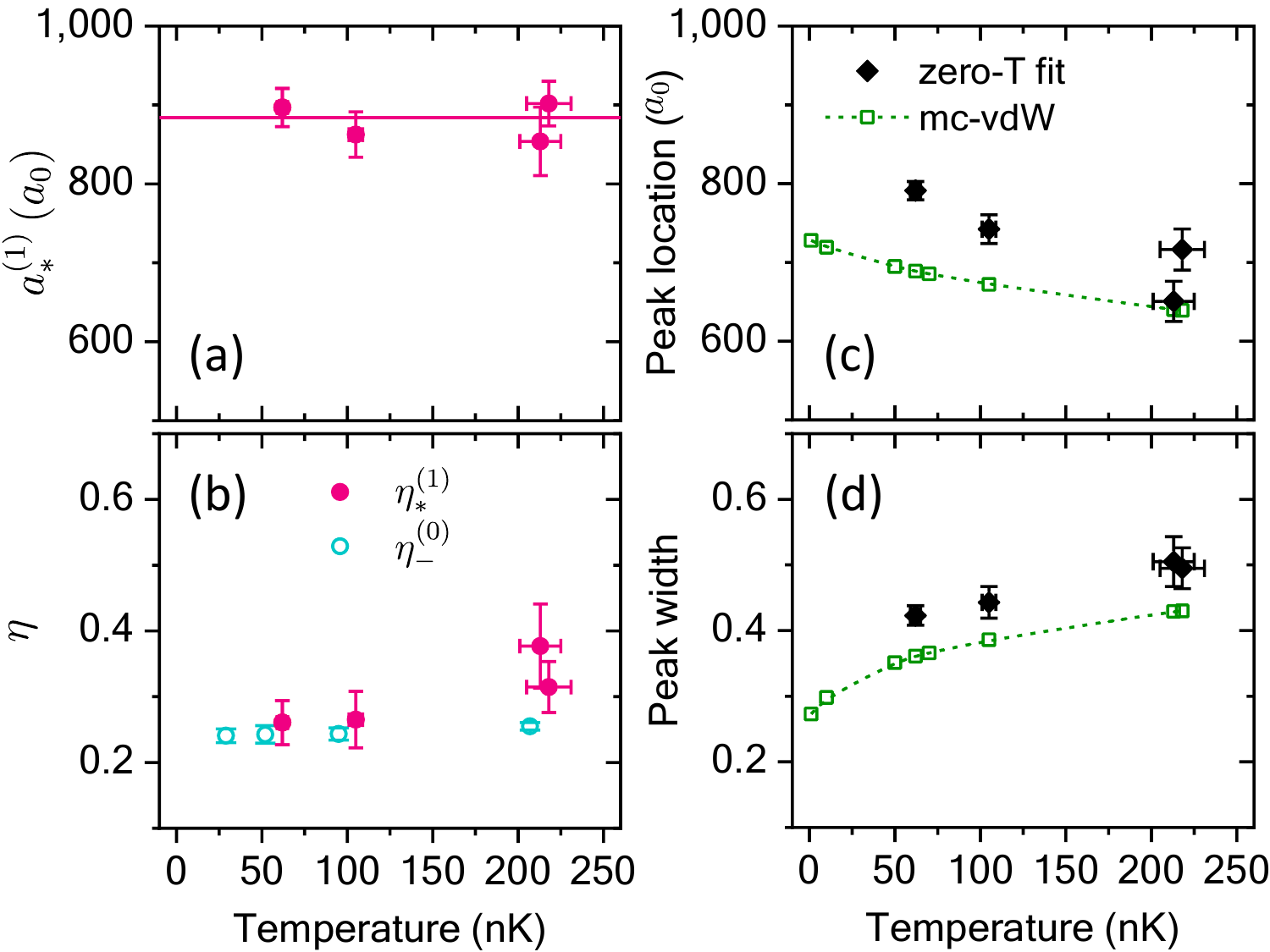}
\centering
\caption{Summary of the fit results on atom-dimer resonance and comparison with mc-vdW theory. (a) $a_{*}^{(1)}$ and (b) $\eta_{*}^{(1)}$ extracted from the finite-temperature model fits (magenta circles). The horizontal line indicates the mean value of the four points. $\eta_{*}^{(1)}$ is found to be consistent with $\eta_{-}^{(0)}$ reported in our previous work \cite{chapurin2019prl}. (c) Phenomenological peak location and (d) width extracted from the zero-temperature fits to the finite-temperature data (black diamonds) or model (green squares). Finite temperature effects not only shift the peak location but also greatly broaden the peaks. Both behaviors are captured by our mc-vdW model.
}
\label{fig:astar} 
\end{figure}

The variation of the fitting parameters with temperature is summarized in Fig.~\ref{fig:astar}. We contrast the fit results from the above mentioned finite-temperature model [panel (a) and (b)] \cite{braaten2004pra} and from a zero-temperature model [panel (c) and (d)]. The former model reveals an energy independent parameter $a_{*}^{(1)}$ that is approached by the phenomenological peak location from the latter model as $T \to 0$. We determine $a_{*}^{(1)} = 884(14) \thinspace a_0 = 13.7(2) \thinspace r_\mathrm{vdW}$ from the weighted mean of the four experimental points [Fig.~\ref{fig:astar}(a)]. Notably, the inelasticity parameter $\eta_{*}^{(1)} = 0.28(2)$ overlaps with the previously measured $\eta_{-}^{(0)} = 0.25(1)$ for $a<0$ within uncertainty [Fig.~\ref{fig:astar}(b)], consistent with the expected continuity of Efimov physics across a two-body resonance.

\begin{figure}
\includegraphics[width=8.5cm,keepaspectratio]{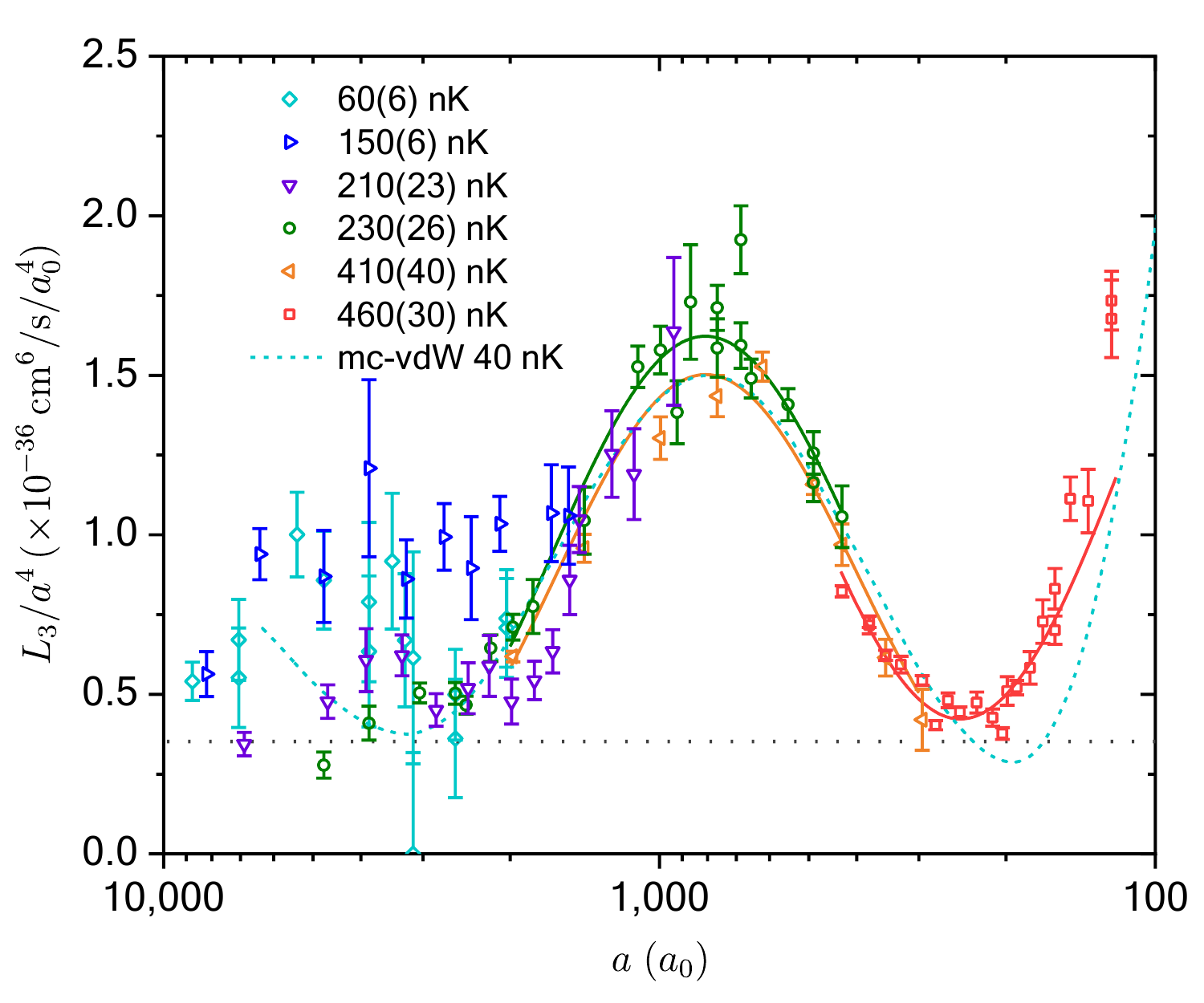}
\centering
\caption{Three-body recombination coefficient $L_{3}$ divided by $a^4$ as a function of $a$. Colored solid lines represent fits to a finite-temperature model either in the neighborhood of the maximum or the minimum. The dotted horizontal line represents three-body recombination into deeply bound dimers with $\eta_{+}^{(0)} = 0.20$. The cyan dashed line represents our mc-vdW calculation done for $40 \thinspace \mathrm{nK}$ \cite{SuppMat}.
}
\label{fig:recombination} 
\end{figure}

%% AAA loss
We determine the remaining two Efimov features for $a>0$ through measurements of three-body recombination coefficient $L_3$. Unlike for $a<0$, there are no expected three-body resonances in $L_3$ for $a>0$. Instead, two indistinguishable decay pathways lead to interference minima and maxima \cite{braaten2006pr, greene2017rmp, dincao2018jpb}, denoted as $a_{+}^{(n)}$ and $a_{p}^{(n)}$ respectively. Upon finishing the evaporation, we ramp up the trap depth adiabatically to about 10 times the final temperature to avoid number loss due to ongoing evaporation. The peak value of the phase-space density is always restricted below $1$ to ensure Boltzmann statistics. We use a rate equation \cite{SuppMat} to describe the time evolution of atom number and temperature to obtain the value of $L_3$ at various $a$. Since the overall scaling of $L_3$ is proportional to $a^4$, we divide out this prefactor in Fig.~\ref{fig:recombination} to emphasize the log-periodic modulation due to Efimov physics. 

To extract the minimum $a_{+}^{(0)}$, we fit the data set of $460 \thinspace \mathrm{nK}$ with the finite-temperature model \cite{braaten2008pra} (red solid line in Fig.~\ref{fig:recombination}). There are three free parameters in our fitting function, $a_{+}$ and $\eta_{+}$ accounting for the location of the minimum and the contrast of the oscillation, and an amplitude-scaling factor $A_+$. We obtain $a_{+}^{(0)} = 246(6) \thinspace a_0$, $\eta_{+}^{(0)} = 0.20(2)$. This result agrees with a fit with a zero-range, zero-energy model \cite{braaten2006pr}, suggesting the negligible effects of finite temperature on $a_{+}^{(0)}$. To extract the maximum $a_{p}^{(0)}$, we fix the contrast parameter to $0.20$ in the same finite temperature model and fit the data sets of $410 \thinspace \mathrm{nK}$ and $230 \thinspace \mathrm{nK}$ with the empirical temperatures as inputs. We determine $a_{p}^{(0)} = 876(28) \thinspace a_0$ from the mean value of the two conditions. For all three temperature fits, the mean value of $A_+$ is within 17\% of unity \cite{SuppMat}, consistent with our density calibration uncertainty of $<10\%$. As we scan $a$ to larger values, poorly-understood temperature and density effects complicate the interpretation of our $L_3$ measurements \cite{SuppMat}.

%% figure 4

%% figure 5
\begin{figure}
\includegraphics[width=8.5cm,keepaspectratio]{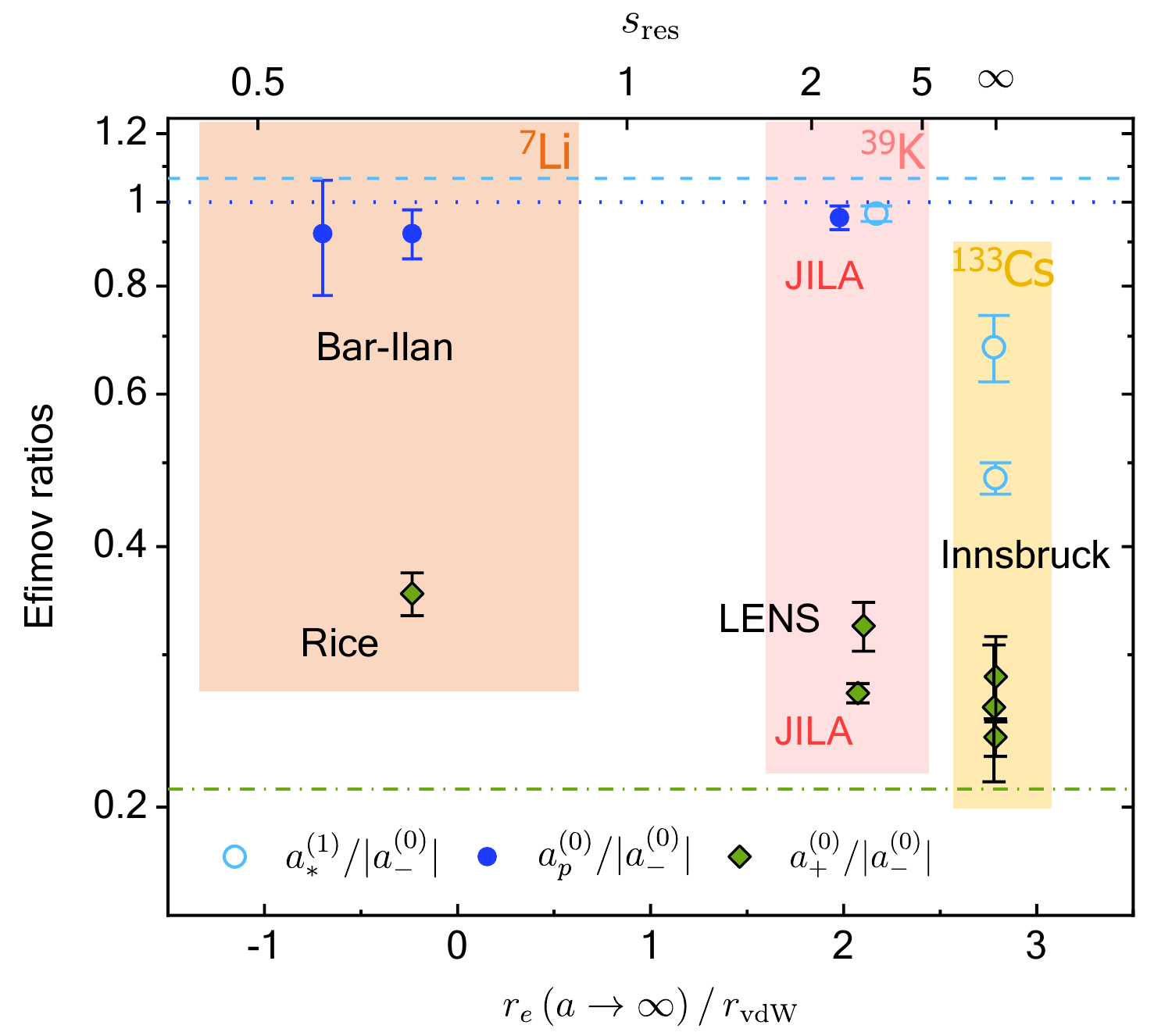} 
\centering
\caption{Summary of experimental \cite{zenesini2014pra, kraemer2006nature, berninger2011prl, ferlaino2011fbs, ferlaino2011fbs, gross2009prl, gross2010prl, dyke2013pra, zaccanti2009natphys, chapurin2019prl} and theoretical results \cite{dincao2018jpb} of three Efimov ratios between features on opposite sides of the Feshbach resonance. Here we show only the experiments on the observables presented in Fig.~\ref{fig:cartoon}. The two JILA points at the top are artificially spaced in horizontal direction for visibility. The corresponding zero-range theory predictions are shown as the dashed, dotted and dash-dotted lines. $r_e$ is evaluated at unitarity ($a \to \infty$) using the model given in \cite{chapurin2019prl}, in which $r_e$ is related to the coupling-strength parameter $s_\mathrm{res}$ \cite{chin2010rmp, chapurin2019prl} that defines strong ($s_\mathrm{res} \gg 1$) and weak ($s_\mathrm{res} \ll 1$) Feshbach resonances.
}
\label{fig:ratio} 
\end{figure}

%% table 1
\begin{table*}
    \begin{center}
\begin{ruledtabular}
\begin{tabular}{c|cc|ccccc|ccc}
    \toprule
& \multicolumn{2}{c|}{Observables for $a<0$} & \multicolumn{5}{c|}{Observables for $a>0$} & \multicolumn{3}{c}{Efimov ratios} \\ \hline
Model & $a_{-}^{(0)} / a_0$ & $\eta_-^{(0)}$ & $a_{*}^{(1)} / a_0$ & $\eta_{*}^{(1)}$ & $a_{+}^{(0)} / a_0$ & $\eta_{+}^{(0)}$ & $a_{p}^{(0)}$ & $a_{*}^{(1)} / | a_{-}^{(0)} |$ & $a_{+}^{(0)} / |a_{-}^{(0)}|$ & $a_{p}^{(0)} / |a_{-}^{(0)}|$  \\ \hline
Efimov universality \cite{dincao2018jpb} & -- & -- & -- & -- & -- & -- & -- & $1.06522$ & $0.209914$ & $1$ \\
VdW universality \cite{mestrom2017pra} & $-626$ & -- & $213$ & -- & $90$ & -- & -- & $0.340$ & $0.143$ & -- \\
Mc-vdW  & $-846(19)$ & $0.21(1)$ \cite{chapurin2019prl} & $809(1)$ & $0.27(3)$ & $200(1)$ & $0.10(1)$ & $817(1)$ & $0.95(2)$ & $0.236(5)$ & $0.96(2)$ \\
Experimental  & $-908(11)$ & $0.25(1)$ \cite{chapurin2019prl} & $884(14)$ & $0.28(2)$ & $246(6)$ & $0.20(2)$ & $876(28)$ & $0.97(2)$ & $0.271(7)$ & $0.96(3)$ \\
\bottomrule
\end{tabular}
\end{ruledtabular}
\end{center}
\caption{Summary of three-body benchmark features characterizing an Efimov spectrum in $\prescript{39}{}{\mathrm{K}}$. The locations of the features and their ratios are compared to various theoretical models. 
Efimov universality is a zero-range theory that defines the universal scaling of the Efimov structure but leaves undetermined the absolute locations of the various features. VdW (van der Waals) universality treats interactions with a single-channel vdW potential. The model pins down the absolute scale of the structure, and quantifies how low-lying Efimov states can be distorted by the vdW interactions. Mc (multi-channel) vdW, elaborated in \cite{SuppMat}, is our attempt at a more realistic theoretical model for our $\prescript{39}{}{\mathrm{K}}$ resonance.
}
\label{tab:sum}
\end{table*}

\textit{Discussion.} We compare our results on Efimov ratios with previous experimental work \cite{heteronuclear} in Fig.~\ref{fig:ratio}. The particular ratio $a_{*}^{(1)} / a_{-}^{(0)}$ substantially deviates from the zero-range universal value when $r_e$ is large and positive. This is the case for strong Feshbach resonances such as those used in $\prescript{133}{}{\mathrm{Cs}}$ \cite{zenesini2014pra}. While for our intermediate-strength resonance in $\prescript{39}{}{\mathrm{K}}$, with suppressed $r_e$ and $a_{-}^{(0)} = -14.05(17) \thinspace r_\mathrm{vdW}$ \cite{chapurin2019prl}, we obtain $a_{*}^{(1)} / a_{-}^{(0)} = -0.97(2)$. This is within $9\%$ of the universal ratio of $-1.065$ \cite{gogolin2008prl}. The much better agreement makes sense in light of the greater separation of length scales. Quantitatively, $| a_{-}^{(0)} |$ and $a_{*}^{(1)}$ are only about 2 to 3 times $r_e$ for $\prescript{133}{}{\mathrm{Cs}}$, whereas they are about 7 times $r_e$ for $\prescript{39}{}{\mathrm{K}}$. Similarly, our observed $a_{p}^{(0)}$ value is also well spaced from $r_e$, and we observe $a_{p}^{(0)}/a_{-}^{(0)}$ within 4\% of its universal value of $-1$ \cite{helfrich2010pra}. The other two reported $a_{p}^{(0)}$ values were measured in $\prescript{7}{}{\mathrm{Li}}$ resonances with $r_e \approx 0$ \cite{gross2009prl, gross2010prl}. Their ratios $a_{p}^{(0)} / a_{-}^{(0)} = -0.92(14)$ and $-0.92(6)$, although of lower precision, are also consistent with Efimov universality. Finally, Efimov universality yields $a_{+}^{(0)} / a_{-}^{(0)} = -1/\sqrt{22.7} = -0.210$. Empirical values of $a_{+}^{(0)} / a_{-}^{(0)}$ in $\prescript{7}{}{\mathrm{Li}}$ \cite{dyke2013pra}, $\prescript{133}{}{\mathrm{Cs}}$ \cite{kraemer2006nature, berninger2011prl, ferlaino2011fbs}, earlier work in $\prescript{39}{}{\mathrm{K}}$ \cite{roy2013prl}, and our own $\prescript{39}{}{\mathrm{K}}$ result differ from this prediction by between 15\% and 70\%. Moreover, there is no trend towards improved agreement with lower $\left| r_e \right|$. We note that the value of $r_e$ at such small $a$ can differ significantly from the value of $r_e$ at $a \to \infty$ \cite{Blackley:eff-range:2014}, and in these circumstances the short-range effects must be analyzed on a case-by-case basis. Not reviewed in Fig.~\ref{fig:ratio} is a measurement \cite{huang2014prl} of $a_{-}^{(1)} / a_{-}^{(0)}$ in $\prescript{133}{}{\mathrm{Cs}}$, which was consistent with Efimov universality.

Finally, we compare our experimental results with various theoretical models in Table \ref{tab:sum}. The central results of this Letter are the excellent three-way agreement between the Efimov universality, mc-vdW and experimental values for the ratios $a_{*}^{(1)} / a_{-}^{(0)}$ and $a_{p}^{(0)} / a_{-}^{(0)}$, and the corresponding disagreement with the predictions of van der Waals universality.

In summary, we have tested Efimov universality in unprecedented detail and found that it holds remarkably well, even for a resonance that deviates substantially from van der Waals universality.

The authors thank J. Bohn, D. Petrov, D. Kang, W. Zwerger and R. Schmidt for useful discussions. This work is supported by NSF Phys-1734006, NASA/JPL 1502690, Marsico Research Chair and NIST.

%\bibliography{bib_file}

\begin{thebibliography}{54}%
\makeatletter
\providecommand \@ifxundefined [1]{%
 \@ifx{#1\undefined}
}%
\providecommand \@ifnum [1]{%
 \ifnum #1\expandafter \@firstoftwo
 \else \expandafter \@secondoftwo
 \fi
}%
\providecommand \@ifx [1]{%
 \ifx #1\expandafter \@firstoftwo
 \else \expandafter \@secondoftwo
 \fi
}%
\providecommand \natexlab [1]{#1}%
\providecommand \enquote  [1]{``#1''}%
\providecommand \bibnamefont  [1]{#1}%
\providecommand \bibfnamefont [1]{#1}%
\providecommand \citenamefont [1]{#1}%
\providecommand \href@noop [0]{\@secondoftwo}%
\providecommand \href [0]{\begingroup \@sanitize@url \@href}%
\providecommand \@href[1]{\@@startlink{#1}\@@href}%
\providecommand \@@href[1]{\endgroup#1\@@endlink}%
\providecommand \@sanitize@url [0]{\catcode `\\12\catcode `\$12\catcode
  `\&12\catcode `\#12\catcode `\^12\catcode `\_12\catcode `\%12\relax}%
\providecommand \@@startlink[1]{}%
\providecommand \@@endlink[0]{}%
\providecommand \url  [0]{\begingroup\@sanitize@url \@url }%
\providecommand \@url [1]{\endgroup\@href {#1}{\urlprefix }}%
\providecommand \urlprefix  [0]{URL }%
\providecommand \Eprint [0]{\href }%
\providecommand \doibase [0]{http://dx.doi.org/}%
\providecommand \selectlanguage [0]{\@gobble}%
\providecommand \bibinfo  [0]{\@secondoftwo}%
\providecommand \bibfield  [0]{\@secondoftwo}%
\providecommand \translation [1]{[#1]}%
\providecommand \BibitemOpen [0]{}%
\providecommand \bibitemStop [0]{}%
\providecommand \bibitemNoStop [0]{.\EOS\space}%
\providecommand \EOS [0]{\spacefactor3000\relax}%
\providecommand \BibitemShut  [1]{\csname bibitem#1\endcsname}%
\let\auto@bib@innerbib\@empty
%</preamble>
\bibitem [{\citenamefont {Efimov}(1970)}]{efimov1970plb}%
  \BibitemOpen
  \bibfield  {author} {\bibinfo {author} {\bibfnamefont {V.}~\bibnamefont
  {Efimov}},\ }\href {\doibase https://doi.org/10.1016/0370-2693(70)90349-7}
  {\bibfield  {journal} {\bibinfo  {journal} {Physics Letters B}\ }\textbf
  {\bibinfo {volume} {33}},\ \bibinfo {pages} {563 } (\bibinfo {year}
  {1970})}\BibitemShut {NoStop}%
\bibitem [{\citenamefont {Braaten}\ and\ \citenamefont
  {Hammer}(2006)}]{braaten2006pr}%
  \BibitemOpen
  \bibfield  {author} {\bibinfo {author} {\bibfnamefont {E.}~\bibnamefont
  {Braaten}}\ and\ \bibinfo {author} {\bibfnamefont {H.-W.}\ \bibnamefont
  {Hammer}},\ }\href {\doibase https://doi.org/10.1016/j.physrep.2006.03.001}
  {\bibfield  {journal} {\bibinfo  {journal} {Physics Reports}\ }\textbf
  {\bibinfo {volume} {428}},\ \bibinfo {pages} {259 } (\bibinfo {year}
  {2006})}\BibitemShut {NoStop}%
\bibitem [{\citenamefont {Greene}\ \emph {et~al.}(2017)\citenamefont {Greene},
  \citenamefont {Giannakeas},\ and\ \citenamefont
  {P\'erez-R\'{\i}os}}]{greene2017rmp}%
  \BibitemOpen
  \bibfield  {author} {\bibinfo {author} {\bibfnamefont {C.~H.}\ \bibnamefont
  {Greene}}, \bibinfo {author} {\bibfnamefont {P.}~\bibnamefont {Giannakeas}},
  \ and\ \bibinfo {author} {\bibfnamefont {J.}~\bibnamefont
  {P\'erez-R\'{\i}os}},\ }\href {\doibase 10.1103/RevModPhys.89.035006}
  {\bibfield  {journal} {\bibinfo  {journal} {Rev. Mod. Phys.}\ }\textbf
  {\bibinfo {volume} {89}},\ \bibinfo {pages} {035006} (\bibinfo {year}
  {2017})}\BibitemShut {NoStop}%
\bibitem [{\citenamefont {D'Incao}(2018)}]{dincao2018jpb}%
  \BibitemOpen
  \bibfield  {author} {\bibinfo {author} {\bibfnamefont {J.~P.}\ \bibnamefont
  {D'Incao}},\ }\href {\doibase 10.1088/1361-6455/aaa116} {\bibfield  {journal}
  {\bibinfo  {journal} {Journal of Physics B: Atomic, Molecular and Optical
  Physics}\ }\textbf {\bibinfo {volume} {51}},\ \bibinfo {pages} {043001}
  (\bibinfo {year} {2018})}\BibitemShut {NoStop}%
\bibitem [{het()}]{heteronuclear}%
  \BibitemOpen
  \href@noop {} {}\bibinfo {note} {Considerable experimental
  \cite{barontini2009prl, williams2009prl, lompe2010sci, bloom2013prl,
  pires2014prl, maier2015prl, johansen2017natphys} and theoretical
  \cite{wang2012prlb, ulmanis2016prl, wang2013aamop, helfrich2011jpb} work has
  explored Efimov universality in systems with unequal atomic masses or
  distinguishable spin states, but to keep the complexity of this Letter
  manageable, we do not review that fascinating topic here.}\BibitemShut
  {Stop}%
\bibitem [{\citenamefont {Gogolin}\ \emph {et~al.}(2008)\citenamefont
  {Gogolin}, \citenamefont {Mora},\ and\ \citenamefont
  {Egger}}]{gogolin2008prl}%
  \BibitemOpen
  \bibfield  {author} {\bibinfo {author} {\bibfnamefont {A.~O.}\ \bibnamefont
  {Gogolin}}, \bibinfo {author} {\bibfnamefont {C.}~\bibnamefont {Mora}}, \
  and\ \bibinfo {author} {\bibfnamefont {R.}~\bibnamefont {Egger}},\ }\href
  {\doibase 10.1103/PhysRevLett.100.140404} {\bibfield  {journal} {\bibinfo
  {journal} {Phys. Rev. Lett.}\ }\textbf {\bibinfo {volume} {100}},\ \bibinfo
  {pages} {140404} (\bibinfo {year} {2008})}\BibitemShut {NoStop}%
\bibitem [{\citenamefont {Helfrich}\ \emph {et~al.}(2010)\citenamefont
  {Helfrich}, \citenamefont {Hammer},\ and\ \citenamefont
  {Petrov}}]{helfrich2010pra}%
  \BibitemOpen
  \bibfield  {author} {\bibinfo {author} {\bibfnamefont {K.}~\bibnamefont
  {Helfrich}}, \bibinfo {author} {\bibfnamefont {H.-W.}\ \bibnamefont
  {Hammer}}, \ and\ \bibinfo {author} {\bibfnamefont {D.~S.}\ \bibnamefont
  {Petrov}},\ }\href {\doibase 10.1103/PhysRevA.81.042715} {\bibfield
  {journal} {\bibinfo  {journal} {Phys. Rev. A}\ }\textbf {\bibinfo {volume}
  {81}},\ \bibinfo {pages} {042715} (\bibinfo {year} {2010})}\BibitemShut
  {NoStop}%
\bibitem [{\citenamefont {D'Incao}\ \emph {et~al.}(2009)\citenamefont
  {D'Incao}, \citenamefont {Greene},\ and\ \citenamefont
  {Esry}}]{dincao2009jpb}%
  \BibitemOpen
  \bibfield  {author} {\bibinfo {author} {\bibfnamefont {J.~P.}\ \bibnamefont
  {D'Incao}}, \bibinfo {author} {\bibfnamefont {C.~H.}\ \bibnamefont {Greene}},
  \ and\ \bibinfo {author} {\bibfnamefont {B.~D.}\ \bibnamefont {Esry}},\
  }\href {\doibase 10.1088/0953-4075/42/4/044016} {\bibfield  {journal}
  {\bibinfo  {journal} {Journal of Physics B: Atomic, Molecular and Optical
  Physics}\ }\textbf {\bibinfo {volume} {42}},\ \bibinfo {pages} {044016}
  (\bibinfo {year} {2009})}\BibitemShut {NoStop}%
\bibitem [{\citenamefont {Berninger}\ \emph {et~al.}(2011)\citenamefont
  {Berninger}, \citenamefont {Zenesini}, \citenamefont {Huang}, \citenamefont
  {Harm}, \citenamefont {N\"agerl}, \citenamefont {Ferlaino}, \citenamefont
  {Grimm}, \citenamefont {Julienne},\ and\ \citenamefont
  {Hutson}}]{berninger2011prl}%
  \BibitemOpen
  \bibfield  {author} {\bibinfo {author} {\bibfnamefont {M.}~\bibnamefont
  {Berninger}}, \bibinfo {author} {\bibfnamefont {A.}~\bibnamefont {Zenesini}},
  \bibinfo {author} {\bibfnamefont {B.}~\bibnamefont {Huang}}, \bibinfo
  {author} {\bibfnamefont {W.}~\bibnamefont {Harm}}, \bibinfo {author}
  {\bibfnamefont {H.-C.}\ \bibnamefont {N\"agerl}}, \bibinfo {author}
  {\bibfnamefont {F.}~\bibnamefont {Ferlaino}}, \bibinfo {author}
  {\bibfnamefont {R.}~\bibnamefont {Grimm}}, \bibinfo {author} {\bibfnamefont
  {P.~S.}\ \bibnamefont {Julienne}}, \ and\ \bibinfo {author} {\bibfnamefont
  {J.~M.}\ \bibnamefont {Hutson}},\ }\href {\doibase
  10.1103/PhysRevLett.107.120401} {\bibfield  {journal} {\bibinfo  {journal}
  {Phys. Rev. Lett.}\ }\textbf {\bibinfo {volume} {107}},\ \bibinfo {pages}
  {120401} (\bibinfo {year} {2011})}\BibitemShut {NoStop}%
\bibitem [{\citenamefont {Wild}\ \emph {et~al.}(2012)\citenamefont {Wild},
  \citenamefont {Makotyn}, \citenamefont {Pino}, \citenamefont {Cornell},\ and\
  \citenamefont {Jin}}]{wild2012prl}%
  \BibitemOpen
  \bibfield  {author} {\bibinfo {author} {\bibfnamefont {R.~J.}\ \bibnamefont
  {Wild}}, \bibinfo {author} {\bibfnamefont {P.}~\bibnamefont {Makotyn}},
  \bibinfo {author} {\bibfnamefont {J.~M.}\ \bibnamefont {Pino}}, \bibinfo
  {author} {\bibfnamefont {E.~A.}\ \bibnamefont {Cornell}}, \ and\ \bibinfo
  {author} {\bibfnamefont {D.~S.}\ \bibnamefont {Jin}},\ }\href {\doibase
  10.1103/PhysRevLett.108.145305} {\bibfield  {journal} {\bibinfo  {journal}
  {Phys. Rev. Lett.}\ }\textbf {\bibinfo {volume} {108}},\ \bibinfo {pages}
  {145305} (\bibinfo {year} {2012})}\BibitemShut {NoStop}%
\bibitem [{\citenamefont {Wang}\ \emph
  {et~al.}(2012{\natexlab{a}})\citenamefont {Wang}, \citenamefont {D'Incao},
  \citenamefont {Esry},\ and\ \citenamefont {Greene}}]{wang2012prl}%
  \BibitemOpen
  \bibfield  {author} {\bibinfo {author} {\bibfnamefont {J.}~\bibnamefont
  {Wang}}, \bibinfo {author} {\bibfnamefont {J.~P.}\ \bibnamefont {D'Incao}},
  \bibinfo {author} {\bibfnamefont {B.~D.}\ \bibnamefont {Esry}}, \ and\
  \bibinfo {author} {\bibfnamefont {C.~H.}\ \bibnamefont {Greene}},\ }\href
  {\doibase 10.1103/PhysRevLett.108.263001} {\bibfield  {journal} {\bibinfo
  {journal} {Phys. Rev. Lett.}\ }\textbf {\bibinfo {volume} {108}},\ \bibinfo
  {pages} {263001} (\bibinfo {year} {2012}{\natexlab{a}})}\BibitemShut
  {NoStop}%
\bibitem [{\citenamefont {Schmidt}\ \emph {et~al.}(2012)\citenamefont
  {Schmidt}, \citenamefont {Rath},\ and\ \citenamefont
  {Zwerger}}]{schmidt2012epjb}%
  \BibitemOpen
  \bibfield  {author} {\bibinfo {author} {\bibfnamefont {R.}~\bibnamefont
  {Schmidt}}, \bibinfo {author} {\bibfnamefont {S.}~\bibnamefont {Rath}}, \
  and\ \bibinfo {author} {\bibfnamefont {W.}~\bibnamefont {Zwerger}},\ }\href
  {\doibase 10.1140/epjb/e2012-30841-3} {\bibfield  {journal} {\bibinfo
  {journal} {The European Physical Journal B}\ }\textbf {\bibinfo {volume}
  {85}},\ \bibinfo {pages} {386} (\bibinfo {year} {2012})}\BibitemShut
  {NoStop}%
\bibitem [{\citenamefont {Naidon}\ \emph {et~al.}(2014)\citenamefont {Naidon},
  \citenamefont {Endo},\ and\ \citenamefont {Ueda}}]{naidon2014prl}%
  \BibitemOpen
  \bibfield  {author} {\bibinfo {author} {\bibfnamefont {P.}~\bibnamefont
  {Naidon}}, \bibinfo {author} {\bibfnamefont {S.}~\bibnamefont {Endo}}, \ and\
  \bibinfo {author} {\bibfnamefont {M.}~\bibnamefont {Ueda}},\ }\href {\doibase
  10.1103/PhysRevLett.112.105301} {\bibfield  {journal} {\bibinfo  {journal}
  {Phys. Rev. Lett.}\ }\textbf {\bibinfo {volume} {112}},\ \bibinfo {pages}
  {105301} (\bibinfo {year} {2014})}\BibitemShut {NoStop}%
\bibitem [{\citenamefont {Langmack}\ \emph {et~al.}(2018)\citenamefont
  {Langmack}, \citenamefont {Schmidt},\ and\ \citenamefont
  {Zwerger}}]{langmack2018pra}%
  \BibitemOpen
  \bibfield  {author} {\bibinfo {author} {\bibfnamefont {C.}~\bibnamefont
  {Langmack}}, \bibinfo {author} {\bibfnamefont {R.}~\bibnamefont {Schmidt}}, \
  and\ \bibinfo {author} {\bibfnamefont {W.}~\bibnamefont {Zwerger}},\ }\href
  {\doibase 10.1103/PhysRevA.97.033623} {\bibfield  {journal} {\bibinfo
  {journal} {Phys. Rev. A}\ }\textbf {\bibinfo {volume} {97}},\ \bibinfo
  {pages} {033623} (\bibinfo {year} {2018})}\BibitemShut {NoStop}%
\bibitem [{\citenamefont {Chin}\ \emph {et~al.}(2010)\citenamefont {Chin},
  \citenamefont {Grimm}, \citenamefont {Julienne},\ and\ \citenamefont
  {Tiesinga}}]{chin2010rmp}%
  \BibitemOpen
  \bibfield  {author} {\bibinfo {author} {\bibfnamefont {C.}~\bibnamefont
  {Chin}}, \bibinfo {author} {\bibfnamefont {R.}~\bibnamefont {Grimm}},
  \bibinfo {author} {\bibfnamefont {P.}~\bibnamefont {Julienne}}, \ and\
  \bibinfo {author} {\bibfnamefont {E.}~\bibnamefont {Tiesinga}},\ }\href
  {\doibase https://doi.org/10.1103/RevModPhys.82.1225} {\bibfield  {journal}
  {\bibinfo  {journal} {Rev. Mod. Phys.}\ }\textbf {\bibinfo {volume} {82}},\
  \bibinfo {pages} {1225} (\bibinfo {year} {2010})}\BibitemShut {NoStop}%
\bibitem [{\citenamefont {Taylor}(1972)}]{scatt}%
  \BibitemOpen
  \bibfield  {author} {\bibinfo {author} {\bibfnamefont {J.~R.}\ \bibnamefont
  {Taylor}},\ }\href@noop {} {\emph {\bibinfo {title} {Scattering Theory}}}\
  (\bibinfo  {publisher} {John Wiley and Sons},\ \bibinfo {address} {New
  York},\ \bibinfo {year} {1972})\BibitemShut {NoStop}%
\bibitem [{\citenamefont {Chin}(2011)}]{chin2011arxiv}%
  \BibitemOpen
  \bibfield  {author} {\bibinfo {author} {\bibfnamefont {C.}~\bibnamefont
  {Chin}},\ }\href {https://arxiv.org/abs/1111.1484} {\bibfield  {journal}
  {\bibinfo  {journal} {arXiv:1111.1484}\ } (\bibinfo {year}
  {2011})}\BibitemShut {NoStop}%
\bibitem [{\citenamefont {Gao}(1998)}]{gao1998pra}%
  \BibitemOpen
  \bibfield  {author} {\bibinfo {author} {\bibfnamefont {B.}~\bibnamefont
  {Gao}},\ }\href {\doibase 10.1103/PhysRevA.58.1728} {\bibfield  {journal}
  {\bibinfo  {journal} {Phys. Rev. A}\ }\textbf {\bibinfo {volume} {58}},\
  \bibinfo {pages} {1728} (\bibinfo {year} {1998})}\BibitemShut {NoStop}%
\bibitem [{\citenamefont {Gao}(2011)}]{gao2011pra}%
  \BibitemOpen
  \bibfield  {author} {\bibinfo {author} {\bibfnamefont {B.}~\bibnamefont
  {Gao}},\ }\href {\doibase 10.1103/PhysRevA.84.022706} {\bibfield  {journal}
  {\bibinfo  {journal} {Phys. Rev. A}\ }\textbf {\bibinfo {volume} {84}},\
  \bibinfo {pages} {022706} (\bibinfo {year} {2011})}\BibitemShut {NoStop}%
\bibitem [{\citenamefont {Chapurin}\ \emph {et~al.}(2019)\citenamefont
  {Chapurin}, \citenamefont {Xie}, \citenamefont {Van~de Graaff}, \citenamefont
  {Popowski}, \citenamefont {D'Incao}, \citenamefont {Julienne}, \citenamefont
  {Ye},\ and\ \citenamefont {Cornell}}]{chapurin2019prl}%
  \BibitemOpen
  \bibfield  {author} {\bibinfo {author} {\bibfnamefont {R.}~\bibnamefont
  {Chapurin}}, \bibinfo {author} {\bibfnamefont {X.}~\bibnamefont {Xie}},
  \bibinfo {author} {\bibfnamefont {M.~J.}\ \bibnamefont {Van~de Graaff}},
  \bibinfo {author} {\bibfnamefont {J.~S.}\ \bibnamefont {Popowski}}, \bibinfo
  {author} {\bibfnamefont {J.~P.}\ \bibnamefont {D'Incao}}, \bibinfo {author}
  {\bibfnamefont {P.~S.}\ \bibnamefont {Julienne}}, \bibinfo {author}
  {\bibfnamefont {J.}~\bibnamefont {Ye}}, \ and\ \bibinfo {author}
  {\bibfnamefont {E.~A.}\ \bibnamefont {Cornell}},\ }\href {\doibase
  10.1103/PhysRevLett.123.233402} {\bibfield  {journal} {\bibinfo  {journal}
  {Phys. Rev. Lett.}\ }\textbf {\bibinfo {volume} {123}},\ \bibinfo {pages}
  {233402} (\bibinfo {year} {2019})}\BibitemShut {NoStop}%
\bibitem [{\citenamefont {Wang}\ \emph {et~al.}(2011)\citenamefont {Wang},
  \citenamefont {D'Incao},\ and\ \citenamefont {Greene}}]{wang2011pra}%
  \BibitemOpen
  \bibfield  {author} {\bibinfo {author} {\bibfnamefont {J.}~\bibnamefont
  {Wang}}, \bibinfo {author} {\bibfnamefont {J.~P.}\ \bibnamefont {D'Incao}}, \
  and\ \bibinfo {author} {\bibfnamefont {C.~H.}\ \bibnamefont {Greene}},\
  }\href@noop {} {\bibfield  {journal} {\bibinfo  {journal} {Phys. Rev. A}\
  }\textbf {\bibinfo {volume} {84}},\ \bibinfo {pages} {052721} (\bibinfo
  {year} {2011})}\BibitemShut {NoStop}%
\bibitem [{Sup()}]{SuppMat}%
  \BibitemOpen
  \href@noop {} {}\bibinfo {note} {See Supplemental Material}\BibitemShut
  {NoStop}%
\bibitem [{\citenamefont {D'Incao}\ and\ \citenamefont
  {Julienne}()}]{dincaoinprep}%
  \BibitemOpen
  \bibfield  {author} {\bibinfo {author} {\bibfnamefont {J.~P.}\ \bibnamefont
  {D'Incao}}\ and\ \bibinfo {author} {\bibfnamefont {P.~J.}\ \bibnamefont
  {Julienne}},\ }\href@noop {} {}\bibinfo {note} {In preparation
  (2020)}\BibitemShut {NoStop}%
\bibitem [{\citenamefont {Thompson}\ \emph {et~al.}(2005)\citenamefont
  {Thompson}, \citenamefont {Hodby},\ and\ \citenamefont
  {Wieman}}]{thompson2005prl}%
  \BibitemOpen
  \bibfield  {author} {\bibinfo {author} {\bibfnamefont {S.}~\bibnamefont
  {Thompson}}, \bibinfo {author} {\bibfnamefont {E.}~\bibnamefont {Hodby}}, \
  and\ \bibinfo {author} {\bibfnamefont {C.}~\bibnamefont {Wieman}},\ }\href
  {\doibase 10.1103/PhysRevLett.94.020401} {\bibfield  {journal} {\bibinfo
  {journal} {Physical review letters}\ }\textbf {\bibinfo {volume} {94}},\
  \bibinfo {pages} {020401} (\bibinfo {year} {2005})}\BibitemShut {NoStop}%
\bibitem [{\citenamefont {K\"ohler}\ \emph {et~al.}(2005)\citenamefont
  {K\"ohler}, \citenamefont {Tiesinga},\ and\ \citenamefont
  {Julienne}}]{kohler2005prl}%
  \BibitemOpen
  \bibfield  {author} {\bibinfo {author} {\bibfnamefont {T.}~\bibnamefont
  {K\"ohler}}, \bibinfo {author} {\bibfnamefont {E.}~\bibnamefont {Tiesinga}},
  \ and\ \bibinfo {author} {\bibfnamefont {P.~S.}\ \bibnamefont {Julienne}},\
  }\href {\doibase 10.1103/PhysRevLett.94.020402} {\bibfield  {journal}
  {\bibinfo  {journal} {Phys. Rev. Lett.}\ }\textbf {\bibinfo {volume} {94}},\
  \bibinfo {pages} {020402} (\bibinfo {year} {2005})}\BibitemShut {NoStop}%
\bibitem [{\citenamefont {Hutson}\ and\ \citenamefont
  {Le~Sueur}(2019)}]{molscat:2019}%
  \BibitemOpen
  \bibfield  {author} {\bibinfo {author} {\bibfnamefont {J.~M.}\ \bibnamefont
  {Hutson}}\ and\ \bibinfo {author} {\bibfnamefont {C.~R.}\ \bibnamefont
  {Le~Sueur}},\ }\href {\doibase doi:10.1016/j.cpc.2019.02.014} {\bibfield
  {journal} {\bibinfo  {journal} {Comp. Phys. Comm.}\ }\textbf {\bibinfo
  {volume} {241}},\ \bibinfo {pages} {9} (\bibinfo {year} {2019})}\BibitemShut
  {NoStop}%
\bibitem [{mbf()}]{mbf-github:2020}%
  \BibitemOpen
  \href@noop {} {}\bibinfo {note} {J. M. Hutson and C. R. Le Sueur, ``MOLSCAT,
  BOUND and FIELD, version 2020.0," https://github.com/molscat/molscat
  (2020).}\BibitemShut {Stop}%
\bibitem [{\citenamefont {Frye}\ and\ \citenamefont
  {Hutson}(2020)}]{Frye:quasibound:2020}%
  \BibitemOpen
  \bibfield  {author} {\bibinfo {author} {\bibfnamefont {M.~D.}\ \bibnamefont
  {Frye}}\ and\ \bibinfo {author} {\bibfnamefont {J.~M.}\ \bibnamefont
  {Hutson}},\ }\href@noop {} {\bibfield  {journal} {\bibinfo  {journal} {Phys.
  Rev. Res.}\ }\textbf {\bibinfo {volume} {2}},\ \bibinfo {pages} {013291}
  (\bibinfo {year} {2020})}\BibitemShut {NoStop}%
\bibitem [{\citenamefont {Falke}\ \emph {et~al.}(2008)\citenamefont {Falke},
  \citenamefont {Kn\"ockel}, \citenamefont {Friebe}, \citenamefont {Riedmann},
  \citenamefont {Tiemann},\ and\ \citenamefont {Lisdat}}]{Falke:2008}%
  \BibitemOpen
  \bibfield  {author} {\bibinfo {author} {\bibfnamefont {S.}~\bibnamefont
  {Falke}}, \bibinfo {author} {\bibfnamefont {H.}~\bibnamefont {Kn\"ockel}},
  \bibinfo {author} {\bibfnamefont {J.}~\bibnamefont {Friebe}}, \bibinfo
  {author} {\bibfnamefont {M.}~\bibnamefont {Riedmann}}, \bibinfo {author}
  {\bibfnamefont {E.}~\bibnamefont {Tiemann}}, \ and\ \bibinfo {author}
  {\bibfnamefont {C.}~\bibnamefont {Lisdat}},\ }\href {\doibase
  10.1103/PhysRevA.78.012503} {\bibfield  {journal} {\bibinfo  {journal} {Phys.
  Rev. A}\ }\textbf {\bibinfo {volume} {78}},\ \bibinfo {pages} {012503}
  (\bibinfo {year} {2008})}\BibitemShut {NoStop}%
\bibitem [{\citenamefont {Helfrich}\ and\ \citenamefont
  {Hammer}(2009)}]{helfrich2009epl}%
  \BibitemOpen
  \bibfield  {author} {\bibinfo {author} {\bibfnamefont {K.}~\bibnamefont
  {Helfrich}}\ and\ \bibinfo {author} {\bibfnamefont {H.-W.}\ \bibnamefont
  {Hammer}},\ }\href {\doibase 10.1209/0295-5075/86/53003} {\bibfield
  {journal} {\bibinfo  {journal} {{EPL} (Europhysics Letters)}\ }\textbf
  {\bibinfo {volume} {86}},\ \bibinfo {pages} {53003} (\bibinfo {year}
  {2009})}\BibitemShut {NoStop}%
\bibitem [{\citenamefont {Braaten}\ and\ \citenamefont
  {Hammer}(2004)}]{braaten2004pra}%
  \BibitemOpen
  \bibfield  {author} {\bibinfo {author} {\bibfnamefont {E.}~\bibnamefont
  {Braaten}}\ and\ \bibinfo {author} {\bibfnamefont {H.-W.}\ \bibnamefont
  {Hammer}},\ }\href {\doibase 10.1103/PhysRevA.70.042706} {\bibfield
  {journal} {\bibinfo  {journal} {Phys. Rev. A}\ }\textbf {\bibinfo {volume}
  {70}},\ \bibinfo {pages} {042706} (\bibinfo {year} {2004})}\BibitemShut
  {NoStop}%
\bibitem [{\citenamefont {Braaten}\ \emph {et~al.}(2008)\citenamefont
  {Braaten}, \citenamefont {Hammer}, \citenamefont {Kang},\ and\ \citenamefont
  {Platter}}]{braaten2008pra}%
  \BibitemOpen
  \bibfield  {author} {\bibinfo {author} {\bibfnamefont {E.}~\bibnamefont
  {Braaten}}, \bibinfo {author} {\bibfnamefont {H.-W.}\ \bibnamefont {Hammer}},
  \bibinfo {author} {\bibfnamefont {D.}~\bibnamefont {Kang}}, \ and\ \bibinfo
  {author} {\bibfnamefont {L.}~\bibnamefont {Platter}},\ }\href {\doibase
  10.1103/PhysRevA.78.043605} {\bibfield  {journal} {\bibinfo  {journal} {Phys.
  Rev. A}\ }\textbf {\bibinfo {volume} {78}},\ \bibinfo {pages} {043605}
  (\bibinfo {year} {2008})}\BibitemShut {NoStop}%
\bibitem [{\citenamefont {Zenesini}\ \emph {et~al.}(2014)\citenamefont
  {Zenesini}, \citenamefont {Huang}, \citenamefont {Berninger}, \citenamefont
  {N\"agerl}, \citenamefont {Ferlaino},\ and\ \citenamefont
  {Grimm}}]{zenesini2014pra}%
  \BibitemOpen
  \bibfield  {author} {\bibinfo {author} {\bibfnamefont {A.}~\bibnamefont
  {Zenesini}}, \bibinfo {author} {\bibfnamefont {B.}~\bibnamefont {Huang}},
  \bibinfo {author} {\bibfnamefont {M.}~\bibnamefont {Berninger}}, \bibinfo
  {author} {\bibfnamefont {H.-C.}\ \bibnamefont {N\"agerl}}, \bibinfo {author}
  {\bibfnamefont {F.}~\bibnamefont {Ferlaino}}, \ and\ \bibinfo {author}
  {\bibfnamefont {R.}~\bibnamefont {Grimm}},\ }\href {\doibase
  10.1103/PhysRevA.90.022704} {\bibfield  {journal} {\bibinfo  {journal} {Phys.
  Rev. A}\ }\textbf {\bibinfo {volume} {90}},\ \bibinfo {pages} {022704}
  (\bibinfo {year} {2014})}\BibitemShut {NoStop}%
\bibitem [{\citenamefont {Kraemer}\ \emph {et~al.}(2006)\citenamefont
  {Kraemer}, \citenamefont {Mark}, \citenamefont {Waldburger}, \citenamefont
  {Danzl}, \citenamefont {Chin}, \citenamefont {Engeser}, \citenamefont
  {Lange}, \citenamefont {Pilch}, \citenamefont {Jaakkola}, \citenamefont
  {N\"agerl},\ and\ \citenamefont {Grimm}}]{kraemer2006nature}%
  \BibitemOpen
  \bibfield  {author} {\bibinfo {author} {\bibfnamefont {T.}~\bibnamefont
  {Kraemer}}, \bibinfo {author} {\bibfnamefont {M.}~\bibnamefont {Mark}},
  \bibinfo {author} {\bibfnamefont {P.}~\bibnamefont {Waldburger}}, \bibinfo
  {author} {\bibfnamefont {J.~G.}\ \bibnamefont {Danzl}}, \bibinfo {author}
  {\bibfnamefont {C.}~\bibnamefont {Chin}}, \bibinfo {author} {\bibfnamefont
  {B.}~\bibnamefont {Engeser}}, \bibinfo {author} {\bibfnamefont {A.~D.}\
  \bibnamefont {Lange}}, \bibinfo {author} {\bibfnamefont {K.}~\bibnamefont
  {Pilch}}, \bibinfo {author} {\bibfnamefont {A.}~\bibnamefont {Jaakkola}},
  \bibinfo {author} {\bibfnamefont {H.-C.}\ \bibnamefont {N\"agerl}}, \ and\
  \bibinfo {author} {\bibfnamefont {R.}~\bibnamefont {Grimm}},\ }\href
  {https://doi.org/10.1038/nature04626} {\bibfield  {journal} {\bibinfo
  {journal} {Nature}\ }\textbf {\bibinfo {volume} {440}},\ \bibinfo {pages}
  {315} (\bibinfo {year} {2006})}\BibitemShut {NoStop}%
\bibitem [{\citenamefont {Ferlaino}\ \emph {et~al.}(2011)\citenamefont
  {Ferlaino}, \citenamefont {Zenesini}, \citenamefont {Berninger},
  \citenamefont {Huang}, \citenamefont {N{\"a}gerl},\ and\ \citenamefont
  {Grimm}}]{ferlaino2011fbs}%
  \BibitemOpen
  \bibfield  {author} {\bibinfo {author} {\bibfnamefont {F.}~\bibnamefont
  {Ferlaino}}, \bibinfo {author} {\bibfnamefont {A.}~\bibnamefont {Zenesini}},
  \bibinfo {author} {\bibfnamefont {M.}~\bibnamefont {Berninger}}, \bibinfo
  {author} {\bibfnamefont {B.}~\bibnamefont {Huang}}, \bibinfo {author}
  {\bibfnamefont {H.-C.}\ \bibnamefont {N{\"a}gerl}}, \ and\ \bibinfo {author}
  {\bibfnamefont {R.}~\bibnamefont {Grimm}},\ }\href {\doibase
  10.1007/s00601-011-0260-7} {\bibfield  {journal} {\bibinfo  {journal}
  {Few-Body Systems}\ }\textbf {\bibinfo {volume} {51}},\ \bibinfo {pages}
  {113} (\bibinfo {year} {2011})}\BibitemShut {NoStop}%
\bibitem [{\citenamefont {Gross}\ \emph {et~al.}(2009)\citenamefont {Gross},
  \citenamefont {Shotan}, \citenamefont {Kokkelmans},\ and\ \citenamefont
  {Khaykovich}}]{gross2009prl}%
  \BibitemOpen
  \bibfield  {author} {\bibinfo {author} {\bibfnamefont {N.}~\bibnamefont
  {Gross}}, \bibinfo {author} {\bibfnamefont {Z.}~\bibnamefont {Shotan}},
  \bibinfo {author} {\bibfnamefont {S.}~\bibnamefont {Kokkelmans}}, \ and\
  \bibinfo {author} {\bibfnamefont {L.}~\bibnamefont {Khaykovich}},\ }\href
  {\doibase 10.1103/PhysRevLett.103.163202} {\bibfield  {journal} {\bibinfo
  {journal} {Phys. Rev. Lett.}\ }\textbf {\bibinfo {volume} {103}},\ \bibinfo
  {pages} {163202} (\bibinfo {year} {2009})}\BibitemShut {NoStop}%
\bibitem [{\citenamefont {Gross}\ \emph {et~al.}(2010)\citenamefont {Gross},
  \citenamefont {Shotan}, \citenamefont {Kokkelmans},\ and\ \citenamefont
  {Khaykovich}}]{gross2010prl}%
  \BibitemOpen
  \bibfield  {author} {\bibinfo {author} {\bibfnamefont {N.}~\bibnamefont
  {Gross}}, \bibinfo {author} {\bibfnamefont {Z.}~\bibnamefont {Shotan}},
  \bibinfo {author} {\bibfnamefont {S.}~\bibnamefont {Kokkelmans}}, \ and\
  \bibinfo {author} {\bibfnamefont {L.}~\bibnamefont {Khaykovich}},\ }\href
  {\doibase 10.1103/PhysRevLett.105.103203} {\bibfield  {journal} {\bibinfo
  {journal} {Phys. Rev. Lett.}\ }\textbf {\bibinfo {volume} {105}},\ \bibinfo
  {pages} {103203} (\bibinfo {year} {2010})}\BibitemShut {NoStop}%
\bibitem [{\citenamefont {Dyke}\ \emph {et~al.}(2013)\citenamefont {Dyke},
  \citenamefont {Pollack},\ and\ \citenamefont {Hulet}}]{dyke2013pra}%
  \BibitemOpen
  \bibfield  {author} {\bibinfo {author} {\bibfnamefont {P.}~\bibnamefont
  {Dyke}}, \bibinfo {author} {\bibfnamefont {S.~E.}\ \bibnamefont {Pollack}}, \
  and\ \bibinfo {author} {\bibfnamefont {R.~G.}\ \bibnamefont {Hulet}},\ }\href
  {\doibase 10.1103/PhysRevA.88.023625} {\bibfield  {journal} {\bibinfo
  {journal} {Phys. Rev. A}\ }\textbf {\bibinfo {volume} {88}},\ \bibinfo
  {pages} {023625} (\bibinfo {year} {2013})}\BibitemShut {NoStop}%
\bibitem [{\citenamefont {Zaccanti}\ \emph {et~al.}(2009)\citenamefont
  {Zaccanti}, \citenamefont {Deissler}, \citenamefont {D'Errico},
  \citenamefont {Fattori}, \citenamefont {Jona-Lasinio}, \citenamefont
  {M\"uller}, \citenamefont {Roati}, \citenamefont {Inguscio},\ and\
  \citenamefont {Modugno}}]{zaccanti2009natphys}%
  \BibitemOpen
  \bibfield  {author} {\bibinfo {author} {\bibfnamefont {M.}~\bibnamefont
  {Zaccanti}}, \bibinfo {author} {\bibfnamefont {B.}~\bibnamefont {Deissler}},
  \bibinfo {author} {\bibfnamefont {C.}~\bibnamefont {D'Errico}}, \bibinfo
  {author} {\bibfnamefont {M.}~\bibnamefont {Fattori}}, \bibinfo {author}
  {\bibfnamefont {M.}~\bibnamefont {Jona-Lasinio}}, \bibinfo {author}
  {\bibfnamefont {S.}~\bibnamefont {M\"uller}}, \bibinfo {author}
  {\bibfnamefont {G.}~\bibnamefont {Roati}}, \bibinfo {author} {\bibfnamefont
  {M.}~\bibnamefont {Inguscio}}, \ and\ \bibinfo {author} {\bibfnamefont
  {G.}~\bibnamefont {Modugno}},\ }\href {\doibase 10.1038/nphys1334} {\bibfield
   {journal} {\bibinfo  {journal} {Nature Physics}\ }\textbf {\bibinfo {volume}
  {5}},\ \bibinfo {pages} {586} (\bibinfo {year} {2009})}\BibitemShut {NoStop}%
\bibitem [{\citenamefont {Mestrom}\ \emph {et~al.}(2017)\citenamefont
  {Mestrom}, \citenamefont {Wang}, \citenamefont {Greene},\ and\ \citenamefont
  {D'Incao}}]{mestrom2017pra}%
  \BibitemOpen
  \bibfield  {author} {\bibinfo {author} {\bibfnamefont {P.~M.~A.}\
  \bibnamefont {Mestrom}}, \bibinfo {author} {\bibfnamefont {J.}~\bibnamefont
  {Wang}}, \bibinfo {author} {\bibfnamefont {C.~H.}\ \bibnamefont {Greene}}, \
  and\ \bibinfo {author} {\bibfnamefont {J.~P.}\ \bibnamefont {D'Incao}},\
  }\href {\doibase 10.1103/PhysRevA.95.032707} {\bibfield  {journal} {\bibinfo
  {journal} {Phys. Rev. A}\ }\textbf {\bibinfo {volume} {95}},\ \bibinfo
  {pages} {032707} (\bibinfo {year} {2017})}\BibitemShut {NoStop}%
\bibitem [{\citenamefont {Roy}\ \emph {et~al.}(2013)\citenamefont {Roy},
  \citenamefont {Landini}, \citenamefont {Trenkwalder}, \citenamefont
  {Semeghini}, \citenamefont {Spagnolli}, \citenamefont {Simoni}, \citenamefont
  {Fattori}, \citenamefont {Inguscio},\ and\ \citenamefont
  {Modugno}}]{roy2013prl}%
  \BibitemOpen
  \bibfield  {author} {\bibinfo {author} {\bibfnamefont {S.}~\bibnamefont
  {Roy}}, \bibinfo {author} {\bibfnamefont {M.}~\bibnamefont {Landini}},
  \bibinfo {author} {\bibfnamefont {A.}~\bibnamefont {Trenkwalder}}, \bibinfo
  {author} {\bibfnamefont {G.}~\bibnamefont {Semeghini}}, \bibinfo {author}
  {\bibfnamefont {G.}~\bibnamefont {Spagnolli}}, \bibinfo {author}
  {\bibfnamefont {A.}~\bibnamefont {Simoni}}, \bibinfo {author} {\bibfnamefont
  {M.}~\bibnamefont {Fattori}}, \bibinfo {author} {\bibfnamefont
  {M.}~\bibnamefont {Inguscio}}, \ and\ \bibinfo {author} {\bibfnamefont
  {G.}~\bibnamefont {Modugno}},\ }\href {\doibase
  https://doi.org/10.1103/PhysRevLett.111.053202} {\bibfield  {journal}
  {\bibinfo  {journal} {Phys. Rev. Lett.}\ }\textbf {\bibinfo {volume} {111}},\
  \bibinfo {pages} {053202} (\bibinfo {year} {2013})}\BibitemShut {NoStop}%
\bibitem [{\citenamefont {Blackley}\ \emph {et~al.}(2014)\citenamefont
  {Blackley}, \citenamefont {Julienne},\ and\ \citenamefont
  {Hutson}}]{Blackley:eff-range:2014}%
  \BibitemOpen
  \bibfield  {author} {\bibinfo {author} {\bibfnamefont {C.~L.}\ \bibnamefont
  {Blackley}}, \bibinfo {author} {\bibfnamefont {P.~S.}\ \bibnamefont
  {Julienne}}, \ and\ \bibinfo {author} {\bibfnamefont {J.~M.}\ \bibnamefont
  {Hutson}},\ }\href {\doibase 10.1103/PhysRevA.89.042701} {\bibfield
  {journal} {\bibinfo  {journal} {Phys. Rev. A}\ }\textbf {\bibinfo {volume}
  {89}},\ \bibinfo {pages} {042701} (\bibinfo {year} {2014})}\BibitemShut
  {NoStop}%
\bibitem [{\citenamefont {Huang}\ \emph {et~al.}(2014)\citenamefont {Huang},
  \citenamefont {Sidorenkov}, \citenamefont {Grimm},\ and\ \citenamefont
  {Hutson}}]{huang2014prl}%
  \BibitemOpen
  \bibfield  {author} {\bibinfo {author} {\bibfnamefont {B.}~\bibnamefont
  {Huang}}, \bibinfo {author} {\bibfnamefont {L.~A.}\ \bibnamefont
  {Sidorenkov}}, \bibinfo {author} {\bibfnamefont {R.}~\bibnamefont {Grimm}}, \
  and\ \bibinfo {author} {\bibfnamefont {J.~M.}\ \bibnamefont {Hutson}},\
  }\href {\doibase 10.1103/PhysRevLett.112.190401} {\bibfield  {journal}
  {\bibinfo  {journal} {Phys. Rev. Lett.}\ }\textbf {\bibinfo {volume} {112}},\
  \bibinfo {pages} {190401} (\bibinfo {year} {2014})}\BibitemShut {NoStop}%
\bibitem [{\citenamefont {Barontini}\ \emph {et~al.}(2009)\citenamefont
  {Barontini}, \citenamefont {Weber}, \citenamefont {Rabatti}, \citenamefont
  {Catani}, \citenamefont {Thalhammer}, \citenamefont {Inguscio},\ and\
  \citenamefont {Minardi}}]{barontini2009prl}%
  \BibitemOpen
  \bibfield  {author} {\bibinfo {author} {\bibfnamefont {G.}~\bibnamefont
  {Barontini}}, \bibinfo {author} {\bibfnamefont {C.}~\bibnamefont {Weber}},
  \bibinfo {author} {\bibfnamefont {F.}~\bibnamefont {Rabatti}}, \bibinfo
  {author} {\bibfnamefont {J.}~\bibnamefont {Catani}}, \bibinfo {author}
  {\bibfnamefont {G.}~\bibnamefont {Thalhammer}}, \bibinfo {author}
  {\bibfnamefont {M.}~\bibnamefont {Inguscio}}, \ and\ \bibinfo {author}
  {\bibfnamefont {F.}~\bibnamefont {Minardi}},\ }\href {\doibase
  10.1103/PhysRevLett.103.043201} {\bibfield  {journal} {\bibinfo  {journal}
  {Phys. Rev. Lett.}\ }\textbf {\bibinfo {volume} {103}},\ \bibinfo {pages}
  {043201} (\bibinfo {year} {2009})}\BibitemShut {NoStop}%
\bibitem [{\citenamefont {Williams}\ \emph {et~al.}(2009)\citenamefont
  {Williams}, \citenamefont {Hazlett}, \citenamefont {Huckans}, \citenamefont
  {Stites}, \citenamefont {Zhang},\ and\ \citenamefont
  {O'Hara}}]{williams2009prl}%
  \BibitemOpen
  \bibfield  {author} {\bibinfo {author} {\bibfnamefont {J.~R.}\ \bibnamefont
  {Williams}}, \bibinfo {author} {\bibfnamefont {E.~L.}\ \bibnamefont
  {Hazlett}}, \bibinfo {author} {\bibfnamefont {J.~H.}\ \bibnamefont
  {Huckans}}, \bibinfo {author} {\bibfnamefont {R.~W.}\ \bibnamefont {Stites}},
  \bibinfo {author} {\bibfnamefont {Y.}~\bibnamefont {Zhang}}, \ and\ \bibinfo
  {author} {\bibfnamefont {K.~M.}\ \bibnamefont {O'Hara}},\ }\href {\doibase
  10.1103/PhysRevLett.103.130404} {\bibfield  {journal} {\bibinfo  {journal}
  {Phys. Rev. Lett.}\ }\textbf {\bibinfo {volume} {103}},\ \bibinfo {pages}
  {130404} (\bibinfo {year} {2009})}\BibitemShut {NoStop}%
\bibitem [{\citenamefont {Lompe}\ \emph {et~al.}(2010)\citenamefont {Lompe},
  \citenamefont {Ottenstein}, \citenamefont {Serwane}, \citenamefont {Wenz},
  \citenamefont {Z{\"u}rn},\ and\ \citenamefont {Jochim}}]{lompe2010sci}%
  \BibitemOpen
  \bibfield  {author} {\bibinfo {author} {\bibfnamefont {T.}~\bibnamefont
  {Lompe}}, \bibinfo {author} {\bibfnamefont {T.~B.}\ \bibnamefont
  {Ottenstein}}, \bibinfo {author} {\bibfnamefont {F.}~\bibnamefont {Serwane}},
  \bibinfo {author} {\bibfnamefont {A.~N.}\ \bibnamefont {Wenz}}, \bibinfo
  {author} {\bibfnamefont {G.}~\bibnamefont {Z{\"u}rn}}, \ and\ \bibinfo
  {author} {\bibfnamefont {S.}~\bibnamefont {Jochim}},\ }\href {\doibase
  10.1126/science.1193148} {\bibfield  {journal} {\bibinfo  {journal}
  {Science}\ }\textbf {\bibinfo {volume} {330}},\ \bibinfo {pages} {940}
  (\bibinfo {year} {2010})}\BibitemShut {NoStop}%
\bibitem [{\citenamefont {Bloom}\ \emph {et~al.}(2013)\citenamefont {Bloom},
  \citenamefont {Hu}, \citenamefont {Cumby},\ and\ \citenamefont
  {Jin}}]{bloom2013prl}%
  \BibitemOpen
  \bibfield  {author} {\bibinfo {author} {\bibfnamefont {R.~S.}\ \bibnamefont
  {Bloom}}, \bibinfo {author} {\bibfnamefont {M.-G.}\ \bibnamefont {Hu}},
  \bibinfo {author} {\bibfnamefont {T.~D.}\ \bibnamefont {Cumby}}, \ and\
  \bibinfo {author} {\bibfnamefont {D.~S.}\ \bibnamefont {Jin}},\ }\href
  {\doibase 10.1103/PhysRevLett.111.105301} {\bibfield  {journal} {\bibinfo
  {journal} {Phys. Rev. Lett.}\ }\textbf {\bibinfo {volume} {111}},\ \bibinfo
  {pages} {105301} (\bibinfo {year} {2013})}\BibitemShut {NoStop}%
\bibitem [{\citenamefont {Pires}\ \emph {et~al.}(2014)\citenamefont {Pires},
  \citenamefont {Ulmanis}, \citenamefont {H\"afner}, \citenamefont {Repp},
  \citenamefont {Arias}, \citenamefont {Kuhnle},\ and\ \citenamefont
  {Weidem\"uller}}]{pires2014prl}%
  \BibitemOpen
  \bibfield  {author} {\bibinfo {author} {\bibfnamefont {R.}~\bibnamefont
  {Pires}}, \bibinfo {author} {\bibfnamefont {J.}~\bibnamefont {Ulmanis}},
  \bibinfo {author} {\bibfnamefont {S.}~\bibnamefont {H\"afner}}, \bibinfo
  {author} {\bibfnamefont {M.}~\bibnamefont {Repp}}, \bibinfo {author}
  {\bibfnamefont {A.}~\bibnamefont {Arias}}, \bibinfo {author} {\bibfnamefont
  {E.~D.}\ \bibnamefont {Kuhnle}}, \ and\ \bibinfo {author} {\bibfnamefont
  {M.}~\bibnamefont {Weidem\"uller}},\ }\href {\doibase
  10.1103/PhysRevLett.112.250404} {\bibfield  {journal} {\bibinfo  {journal}
  {Phys. Rev. Lett.}\ }\textbf {\bibinfo {volume} {112}},\ \bibinfo {pages}
  {250404} (\bibinfo {year} {2014})}\BibitemShut {NoStop}%
\bibitem [{\citenamefont {Maier}\ \emph {et~al.}(2015)\citenamefont {Maier},
  \citenamefont {Eisele}, \citenamefont {Tiemann},\ and\ \citenamefont
  {Zimmermann}}]{maier2015prl}%
  \BibitemOpen
  \bibfield  {author} {\bibinfo {author} {\bibfnamefont {R.~A.~W.}\
  \bibnamefont {Maier}}, \bibinfo {author} {\bibfnamefont {M.}~\bibnamefont
  {Eisele}}, \bibinfo {author} {\bibfnamefont {E.}~\bibnamefont {Tiemann}}, \
  and\ \bibinfo {author} {\bibfnamefont {C.}~\bibnamefont {Zimmermann}},\
  }\href {\doibase 10.1103/PhysRevLett.115.043201} {\bibfield  {journal}
  {\bibinfo  {journal} {Phys. Rev. Lett.}\ }\textbf {\bibinfo {volume} {115}},\
  \bibinfo {pages} {043201} (\bibinfo {year} {2015})}\BibitemShut {NoStop}%
\bibitem [{\citenamefont {Johansen}\ \emph {et~al.}(2017)\citenamefont
  {Johansen}, \citenamefont {DeSalvo}, \citenamefont {Patel},\ and\
  \citenamefont {Chin}}]{johansen2017natphys}%
  \BibitemOpen
  \bibfield  {author} {\bibinfo {author} {\bibfnamefont {J.}~\bibnamefont
  {Johansen}}, \bibinfo {author} {\bibfnamefont {B.}~\bibnamefont {DeSalvo}},
  \bibinfo {author} {\bibfnamefont {K.}~\bibnamefont {Patel}}, \ and\ \bibinfo
  {author} {\bibfnamefont {C.}~\bibnamefont {Chin}},\ }\href {\doibase
  https://doi.org/10.1038/nphys4130} {\bibfield  {journal} {\bibinfo  {journal}
  {Nat. Phys.}\ }\textbf {\bibinfo {volume} {13}},\ \bibinfo {pages} {731}
  (\bibinfo {year} {2017})}\BibitemShut {NoStop}%
\bibitem [{\citenamefont {Wang}\ \emph
  {et~al.}(2012{\natexlab{b}})\citenamefont {Wang}, \citenamefont {Wang},
  \citenamefont {D'Incao},\ and\ \citenamefont {Greene}}]{wang2012prlb}%
  \BibitemOpen
  \bibfield  {author} {\bibinfo {author} {\bibfnamefont {Y.}~\bibnamefont
  {Wang}}, \bibinfo {author} {\bibfnamefont {J.}~\bibnamefont {Wang}}, \bibinfo
  {author} {\bibfnamefont {J.~P.}\ \bibnamefont {D'Incao}}, \ and\ \bibinfo
  {author} {\bibfnamefont {C.~H.}\ \bibnamefont {Greene}},\ }\href {\doibase
  10.1103/PhysRevLett.109.243201} {\bibfield  {journal} {\bibinfo  {journal}
  {Phys. Rev. Lett.}\ }\textbf {\bibinfo {volume} {109}},\ \bibinfo {pages}
  {243201} (\bibinfo {year} {2012}{\natexlab{b}})}\BibitemShut {NoStop}%
\bibitem [{\citenamefont {Ulmanis}\ \emph {et~al.}(2016)\citenamefont
  {Ulmanis}, \citenamefont {H\"afner}, \citenamefont {Pires}, \citenamefont
  {Kuhnle}, \citenamefont {Wang}, \citenamefont {Greene},\ and\ \citenamefont
  {Weidem\"uller}}]{ulmanis2016prl}%
  \BibitemOpen
  \bibfield  {author} {\bibinfo {author} {\bibfnamefont {J.}~\bibnamefont
  {Ulmanis}}, \bibinfo {author} {\bibfnamefont {S.}~\bibnamefont {H\"afner}},
  \bibinfo {author} {\bibfnamefont {R.}~\bibnamefont {Pires}}, \bibinfo
  {author} {\bibfnamefont {E.~D.}\ \bibnamefont {Kuhnle}}, \bibinfo {author}
  {\bibfnamefont {Y.}~\bibnamefont {Wang}}, \bibinfo {author} {\bibfnamefont
  {C.~H.}\ \bibnamefont {Greene}}, \ and\ \bibinfo {author} {\bibfnamefont
  {M.}~\bibnamefont {Weidem\"uller}},\ }\href {\doibase
  10.1103/PhysRevLett.117.153201} {\bibfield  {journal} {\bibinfo  {journal}
  {Phys. Rev. Lett.}\ }\textbf {\bibinfo {volume} {117}},\ \bibinfo {pages}
  {153201} (\bibinfo {year} {2016})}\BibitemShut {NoStop}%
\bibitem [{\citenamefont {Wang}\ \emph {et~al.}(2013)\citenamefont {Wang},
  \citenamefont {D'Incao},\ and\ \citenamefont {Esry}}]{wang2013aamop}%
  \BibitemOpen
  \bibfield  {author} {\bibinfo {author} {\bibfnamefont {Y.}~\bibnamefont
  {Wang}}, \bibinfo {author} {\bibfnamefont {J.~P.}\ \bibnamefont {D'Incao}},
  \ and\ \bibinfo {author} {\bibfnamefont {B.~D.}\ \bibnamefont {Esry}},\ }in\
  \href {\doibase https://doi.org/10.1016/B978-0-12-408090-4.00001-3} {\emph
  {\bibinfo {booktitle} {Advances in Atomic, Molecular, and Optical
  Physics}}},\ Vol.~\bibinfo {volume} {62},\ \bibinfo {editor} {edited by\
  \bibinfo {editor} {\bibfnamefont {E.}~\bibnamefont {Arimondo}}, \bibinfo
  {editor} {\bibfnamefont {P.~R.}\ \bibnamefont {Berman}}, \ and\ \bibinfo
  {editor} {\bibfnamefont {C.~C.}\ \bibnamefont {Lin}}}\ (\bibinfo  {publisher}
  {Academic Press},\ \bibinfo {year} {2013})\ pp.\ \bibinfo {pages} {1 --
  115}\BibitemShut {NoStop}%
\bibitem [{\citenamefont {Helfrich}\ and\ \citenamefont
  {Hammer}(2011)}]{helfrich2011jpb}%
  \BibitemOpen
  \bibfield  {author} {\bibinfo {author} {\bibfnamefont {K.}~\bibnamefont
  {Helfrich}}\ and\ \bibinfo {author} {\bibfnamefont {H.}~\bibnamefont
  {Hammer}},\ }\href {\doibase 10.1088/0953-4075/44/21/215301} {\bibfield
  {journal} {\bibinfo  {journal} {Journal of Physics B-atomic Molecular and
  Optical Physics}\ }\textbf {\bibinfo {volume} {44}} (\bibinfo {year}
  {2011}),\ 10.1088/0953-4075/44/21/215301}\BibitemShut {NoStop}%
\end{thebibliography}


\end{document}

\begin{abstract}
\end{abstract}

\title{Supplemental Material: Observation of Efimov Universality across a Non-Universal Feshbach Resonance in $\prescript{39}{}{\mathrm{K}}$}

\author{Xin Xie}
\affiliation{JILA, National Institute of Standards and Technology and the University of Colorado, and Department of Physics, Boulder, Colorado 80309, USA}
\author{Michael J. Van de Graaff}
\affiliation{JILA, National Institute of Standards and Technology and the University of Colorado, and Department of Physics, Boulder, Colorado 80309, USA}
\author{Roman Chapurin}
\affiliation{JILA, National Institute of Standards and Technology and the University of Colorado, and Department of Physics, Boulder, Colorado 80309, USA}
\author{Matthew D. Frye}
\affiliation{Joint Quantum Centre (JQC) Durham-Newcastle, Department of Chemistry, Durham University, South Road, Durham DH1 3LE, United Kingdom}
\author{Jeremy M. Hutson}
\affiliation{Joint Quantum Centre (JQC) Durham-Newcastle, Department of Chemistry, Durham University, South Road, Durham DH1 3LE, United Kingdom}
\author{Jos\'e P. D'Incao}
\affiliation{JILA, National Institute of Standards and Technology and the University of Colorado, and Department of Physics, Boulder, Colorado 80309, USA}
\author{Paul S. Julienne}
\affiliation{Joint Quantum Institute, National Institute of Standards and Technology and the University of Maryland, College Park, MD 20742}
\author{Jun Ye}
\affiliation{JILA, National Institute of Standards and Technology and the University of Colorado, and Department of Physics, Boulder, Colorado 80309, USA}
\author{Eric A. Cornell}
\affiliation{JILA, National Institute of Standards and Technology and the University of Colorado, and Department of Physics, Boulder, Colorado 80309, USA}

\date{\today}
\maketitle

\section{Coupled-channel Calculation for the Lifetime of Quasibound Dimer State and Measurement of Two-body Decay Coefficient $L_2(a)$}
\label{sec:2btheory}

We have carried out coupled-channel quantum scattering calculations to characterize the inelastic scattering from the incoming channel  $\ket{F_1=1,m_{F_1}=-1} \thinspace + \thinspace \ket{F_2=1,m_{F_2}=-1}$ and the decay of the quasibound state below it. We use the general purpose
quantum scattering package \textsc{molscat} \cite{molscat:2019, molscat}. We perform calculations as a function of magnetic field in the vicinity of the Feshbach resonance near $33 \thinspace \mathrm{G}$. We use the singlet and triplet interaction potentials of Falke \textit{et al}. \cite{PhysRevA.78.012503}, with the small modifications described by Chapurin \textit{et al}. \cite{chapurin2019prl} to reproduce the resonance position accurately. We use a fully uncoupled basis set as described in \cite{PhysRevA.87.032517}. We converge on the energies and widths of quasibound states using the automated procedure of Frye and Hutson \cite{Frye:quasibound:2020}.

The incoming scattering channel has orbital angular momentum $L=0$ for the relative motion of the two atoms. The bound state responsible for the resonance is also of predominantly $L=0$ character (an $s$-wave state). However, there is weak coupling from $L=0$ to channels with $L=2$, which is entirely responsible for inelastic scattering at this threshold and the decay of the quasibound state observed here. The coupling operator is
\begin{equation}
\hat{V}^\mathrm{d}(R) = \lambda(R) \left[\hat{s}_1 \cdot \hat{s}_2 - 3(\hat{s}_1 \cdot \vec e_R)(\hat{s}_2 \cdot \vec e_R) \right],
\label{eqn:coupling}
\end{equation}
where $\hat{s}_1$ and $\hat{s}_2$ are the electron spin operators for the two atoms, and $\vec e_R$ is a unit vector along the internuclear vector of length $R$, and
\begin{equation}
\lambda(R) = E_\mathrm{h} \alpha^2 \left[ A_\mathrm{SO} e^{-\gamma R/a_0} + \frac{g_S^2a_0^3}{4 R^3} \right],
\label{eqn:amplitude}
\end{equation}
where $\alpha$ is the fine structure constant, $g_S$ is the $g$-factor of electron, $E_\mathrm{h}$ is the Hartree energy. The second term in the above equation is the magnetic dipolar interaction between the two electronic spins. The first term represents the second-order spin--orbit interaction \cite{Mies1996}, which is known to be important for heavy alkali-metal atoms but has previously been neglected in calculations on light species such as K, Na and Li.

\begin{figure*}
\includegraphics[width=16.5cm,keepaspectratio]{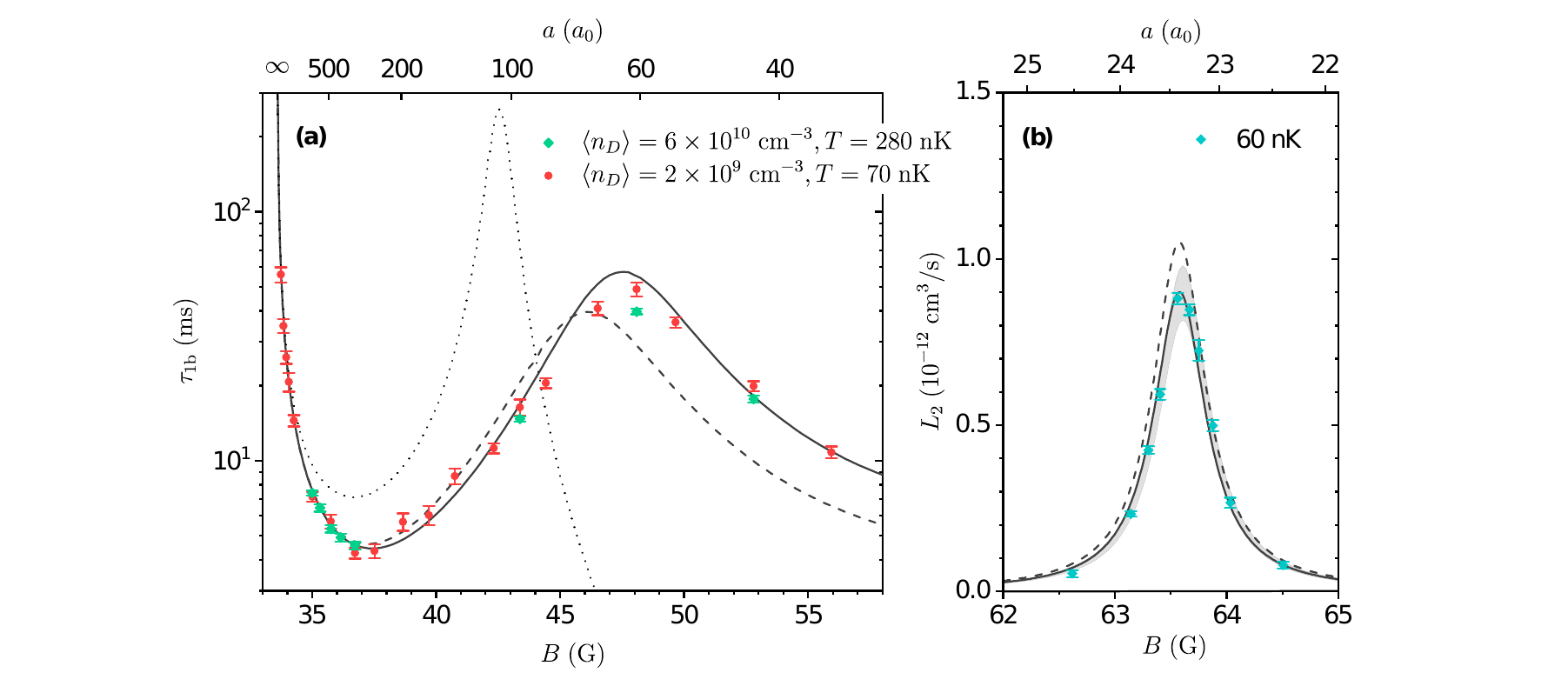} 
\centering
\caption{Refinement of the $d$-wave sector of the couple-channel model for $\prescript{39}{}{\mathrm{K}}$ Feshbach resonance. (a) Lifetime of the quasibound dimer state measured with two different initial conditions. Solid line: calculation including second-order spin--orbit coupling; dashed line: calculation excluding second-order spin--orbit coupling; dotted line: analytic model from \cite{kohler2005prl}. (b) Two-body inelastic collision rate coefficient across the $d$-wave resonance near $64 \thinspace \mathrm{G}$. Gray region indicates the experimental systematic uncertainty on atomic density. Solid and dashed curves are the same as for (a).
}
\label{fig:twobody} 
\end{figure*}

In Fig.~\ref{fig:twobody}(a), the lifetime of the quasibound state, $\tau_\mathrm{1b}$, obtained from coupled-channel calculations is compared with our experimental results. The calculation (dashed line) which neglects second-order spin--orbit coupling shows qualitatively the correct behavior, with a strong peak in the lifetime close to the pole of the resonance and a weaker peak around $46 \thinspace \mathrm{G}$. The peak close to the resonance occurs because in this region the quasibound state is dominated by the component in the incoming channel, whose spatial extent scales as $a$. The subsidiary peak occurs because of interference between the decay amplitudes from the closed-channel and incoming components.

%% figure S1

The position of the subsidiary peak depends on the ratio of the decay amplitudes from the closed-channel and incoming components of the quasibound state. The closed-channel component is concentrated at shorter range than the incoming component. In order to vary the ratio, we therefore introduce a short-range term in the coupling, represented by $A_\mathrm{SO}$ in Eqn.~\ref{eqn:coupling}. The results are insensitive to the exponent $\gamma$, so we fix it at the value 0.7196 obtained for Rb \cite{PhysRevA.82.052514}. The solid line in Fig.~\ref{fig:twobody}(a) shows the lifetime with the value of $A_\mathrm{SO}$ fitted to the experimental lifetimes. The resulting value, $A_\mathrm{SO} = -0.8$, gives a second-order spin--orbit interaction of $-0.015 \, \mathrm{cm}^{-1}$ at the zero-energy inner turning point of the triplet state of $\mathrm{K}_2$, $\sigma_0 = 9.00 \thinspace a_0$. This may be compared with the value for Rb$_2$ \cite{PhysRevA.82.052514}, which is $-0.11$ cm$^{-1}$ at its inner turning point, $\sigma_0 = 9.54 \thinspace a_0$. The ratio about 7 of the values for Rb and K is a little less than half the ratio of the squares of the spin--orbit splittings in the lowest atomic $\prescript{2}{}{\mathrm{P}}$ states.

K\"ohler \textit{et al}. \cite{kohler2005prl} developed a model for the lifetime of the quasibound state near a resonance and obtained the expression
\begin{equation}
\tau_\mathrm{1b} = \frac{4 \pi a^3}{L_2(a)},
\end{equation}
where $L_2$ is the zero-energy rate coefficient for two-body inelastic collision from the incoming channel. This result is shown as a dotted line in Fig.~\ref{fig:twobody}(a), using the values of $a$ and $L_2$ obtained from the coupled-channel calculation including second-order spin--orbit coupling. It may be seen that the model is accurate close to the resonance and produces a subsidiary peak in the lifetime further away, but the subsidiary peak is far too high and its position is not correct. This deviation arises because the simple model accounts incorrectly for the interference between the decay amplitudes arising from the closed-channel and incoming components of the quasibound state, which is different for $\tau_\mathrm{1b}$ and $L_2$.

The second-order spin--orbit interaction partially cancels the magnetic dipolar coupling. For the optimized value of $A_\mathrm{SO}$, the coupling function $\lambda(R)$ crosses zero inside the classically forbidden region for the triplet state. In $\prescript{39}{}{\mathrm{K}}$ the overall coupling is thus dominated by the magnetic dipolar term, and the effect of the second-order spin--orbit term is to reduce slightly the total decay rates due to spin relaxation. To provide an independent sanity check on the accuracy of $\lambda(R)$, we perform measurements on the two-body loss coefficient $L_2$ across the narrow $d$-wave resonance near $63.6 \thinspace \mathrm{G}$. The experimental procedure is much the same as that used to determine $L_3$ ($a>0$) as described in the main text. The resonance is determined to be centered at $63.609(4) \thinspace \mathrm{G}$ with full-width-at-half-maximum of $0.567(12) \thinspace \mathrm{G}$. The results are shown in Fig.~\ref{fig:twobody}(b). There is a small discrepancy in the calculated position of the resonance, due to remaining deficiencies in the singlet and/or triplet potential curves. Without $A_\mathrm{SO}$ the prediction (dashed line) overestimates the height of the peak, whereas the prediction including $A_\mathrm{SO}$ agrees with the height and width within the experimental uncertainties. The overall consistency between theory and experiment in the two-body sector validates the reliability of our detection scheme for both molecules and free atoms.

\section{Three-body Coupled-Channel Model}
\label{sec:3btheory}

In the present study we use the same three-body theoretical model for $^{39}$K that we developed in \cite{chapurin2019prl}, using the the adiabatic hyperspherical representation \cite{suno2002pra, wang2011pra, wang2012prl, wang2012PRLb, mestrom2017pra}. A more detailed description of our model can be found in \cite{dincaoinprep}. Briefly, our model includes the proper atomic hyperfine structure via Feshbach projectors in an approach similar to the one used in \cite{jonsell2004ipb}. We replace the actual singlet and triplet potentials by two Lennard-Jones potentials, $v_{S}(r)=-C_6/r^6(1-\lambda_\mathrm{S}^6/r^6)$, and $v_{T}(r)=-C_6/r^6(1-\lambda_{T}^6/r^6)$, with $\lambda_{S}$ and $\lambda_{T}$ adjusted to correctly produce the singlet and triplet scattering lengths but with a much reduced number of bound states than the real interactions. In order to obtain an accurate description of the $\prescript{39}{}{\mathrm{K}}$ Feshbach resonance at $B_0 = 33.5820(14) \thinspace \mathrm{G}$, we allowed for small variations in $a_{S}$ and $a_{T}$ to fine tune the background scattering length $a_{\mathrm{bg}}$ and the width of the Feshbach resonance $\Delta B$, thus ensuring the proper pole strength $a_{\rm bg}\Delta{B}$ (or equivalently, the $s_\mathrm{res}$) \cite{chin2010rmp}.

We determine the relevant three-body scattering observables using $v_{S}$ and $v_{T}$ potentials supporting different number of $s$-wave bound states. This is done by solving the hyperradial Schr\"odinger equation 
\cite{wang2011pra},
\begin{align}
&\left[-\frac{\hbar^2}{2\mu}\frac{d^2}{dR^2}+U_{\nu}(R)\right]F_{\nu}(R)\nonumber\\
&~~~~~~+\sum_{\nu'}W_{\nu\nu'}(R)F_{\nu'}(R)=EF_{\nu}(R),
\label{Schro}
\end{align}
where the hyperradius $R$ gives the overall size of the system, $U_{\nu}(R)$ are the three-body adiabatic potentials, $\nu$ is an index that labels all necessary quantum numbers to characterize each channel, and $E$ is the total energy. From the above equation we determine the relevant scattering $S$-matrix for the recombination rate $L_3$ \cite{wang2011pra}, and atom-dimer relaxation rate $\beta_\mathrm{AD}$ \cite{dincao2008pra}. The values for the locations of the Efimov features, as well as the corresponding inelasticity parameters $\eta$, are then determined via fitting to the universal formulas summarized in \cite{braaten2006pr}. The results are listed in Table~\ref{Tab3BP} for different numbers of $s$-wave singlet (triplet) bound states $N_{S}\equiv N$ ($N_{T}=N-1$) that our model potential can support. We also list the total number of bound states in Table~\ref{Tab3BP}, including all partial waves. Similar to what was found in \cite{chapurin2019prl}, our results in Table~\ref{Tab3BP} show a considerable dependence on $N$, however, approaching to a limiting value for large values of $N$. We extrapolate our results using a fitting formula $f(N)=\alpha+\beta/N^5$, with $\alpha$ and $\beta$ serving as our fitting parameters.

\begin{table}
\begin{ruledtabular}
\begin{tabular}{llll}
$N$ & $N_{\mathrm{tot}}$ & $a_{+}^{(0)}/a_{0}$ & $\eta_+^{(0)}$\\ \hline
2 & 7   & $150(2)$ & 0.01(1)\\
3 & 28 & $193(6)$ & 0.09(4) \\
4 & 60 & $199(2)$ & 0.10(1) \\
$\infty$ & $-$ & $200(1)$ & 0.10(1) \\ [0.1in]
Exp. &  & $246(6)$ & 0.20(2) \\ [0.05in] \hline
$N$ & $N_{\mathrm{tot}}$ & $a_{+}^{(1)}/a_{0}$ & $\eta_+^{(1)}$\\ \hline
2 & 7   & $1682(17)$ & 0.07(1)\\
3 & 28 & $2814(35)$ & 0.22(2) \\
4 & 60 & $3161(14)$ & 0.24(1) \\
$\infty$ & $-$ & $3182(72)$ & 0.24(1) \\ [0.1in]
Exp. &  & $-$ & $-$ \\ [0.05in] \hline
$N$ & $N_{\mathrm{tot}}$ & $a_{p}^{(0)}/a_{0}$ & $\eta_p^{(0)}$\\ \hline
2 & 7   & $459(18)$ & 0.0(1)\\
3 & 28 & $775(10)$ & 0.06(3) \\
4 & 60 & $806(1)$ & 0.13(1) \\
$\infty$ & $-$ & $817(1)$ & 0.14(1) \\ [0.1in]
Exp. &  & $876(28)$ & $-$ \\ [0.05in] \hline
$N$ & $N_{\mathrm{tot}}$ & $a_{*}^{(1)}/a_{0}$ & $\eta_*^{(1)}$\\ \hline
2 & 7   & $461(5)$ & 0.16(1)\\
3 & 28 & $771(9)$ & 0.35(1) \\
4 & 60 & $798(1)$ & 0.20(1) \\
$\infty$ & $-$ & $809(1)$ & 0.27(3) \\ [0.1in]
Exp. &  & $884(14)$ & $0.28(2)$ 
\end{tabular}
\end{ruledtabular}
\label{Tab3BP}
\caption{Values for the locations of Efimov features and the corresponding inelasticity parameter $\eta$ as a function of number of $s$-wave singlet bound states $N$ (or the total number of bound states $N_{\mathrm{tot}}$). The last line in each group lists experimental results obtained in the 
present work.}
\end{table}

\section{Measurements of the Quasibound State Lifetime}
\label{sec:AD}

In our experiment of atom-dimer inelastic collision, we generate Feshbach molecules through magnetoassociation. We scan the magnetic field to a negative scattering length and then wait $10 \, \mathrm{ms}$ to damp out the cloud's breathing motion that is excited by the rapid magnetic field change. We then ramp the field across the resonance to about $500 \, a_0$. At this field, we vary (or eliminate) the number density of the residual atoms: we apply a Gaussian-enveloped adiabatic rapid passage (ARP) pulse followed by an optical blasting pulse to eject the atoms either thoroughly or partially. The binding energy of the molecules at this home field is about $500 \, \mathrm{kHz}$, which is large enough for the $\pm150 \, \mathrm{kHz}$ span of the ARP pulse to flip free atoms into $\ket{F=2,m_F=-2}$ state while the molecules remain intact in the original $\ket{F=1,m_F=-1}$ state. We control the temperature of the samples by adjusting the evaporation sequence prior to the molecule association step. After all the preparations, we jump the magnetic field within $0.5 \, \mathrm{ms}$ to the target science field and let the dimers decay for a variable dwell time. In the end, we jump the field back to $500 \, a_0$ for detection.

\begin{figure}
\includegraphics[width=8.3cm,keepaspectratio]{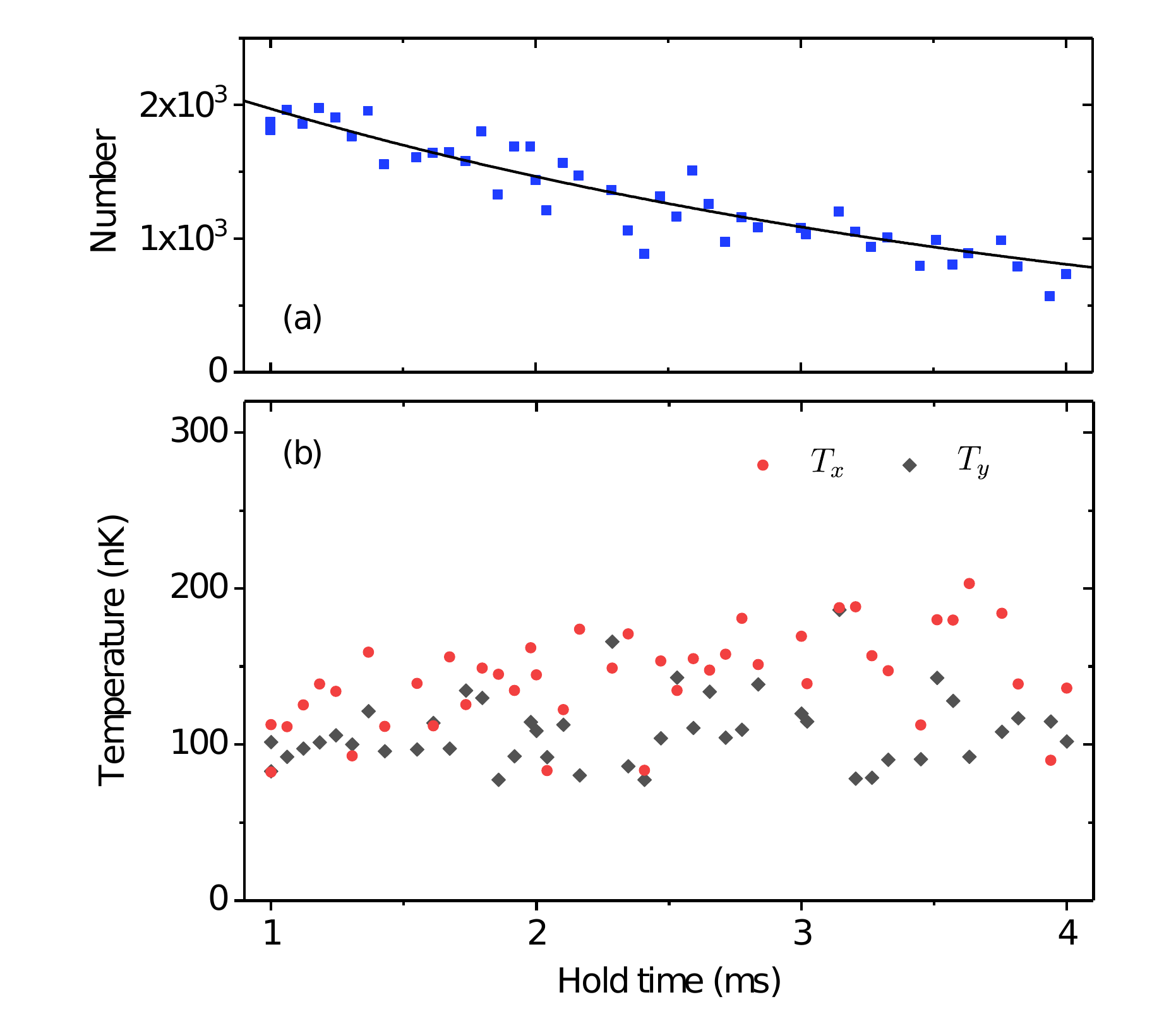} 
\centering
\caption{Time evolution of dimers measured at $1200 \, a_0$ with atoms being the majority in the sample. (a) Dimer number decay (blue squares) fitted with an exponential function (solid line) with a $1/e$ lifetime $\tau_{\mathrm{tot}}$. (b) Temperature of dimers extracted from the horizontal ($\hat{x}$, red circles) and vertical ($\hat{y}$, gray diamonds) widths of the cloud from \textit{in situ} images of the molecules. Heating effect is minimal for these short hold times.
}
\label{fig:lifetime} 
\end{figure}

For the initial conditions $105$ to $218 \thinspace \mathrm{nK}$, we apply a Gaussian-enveloped rf pulse to dissociate the dimers and spin-flip one of the parent atoms into the imaging state. The frequency of the pulse is chosen to be at the empirical peak response of the molecule dissociation spectrum. The rf pulse power is chosen to deplete the molecule population in a short amount of time. Therefore the expansion of the cloud is minimized, helping to achieve the best signal-to-noise on the samples which consist of as few as thousands of molecules. For the lowest temperature $60 \, \mathrm{nK}$ condition, we have sufficient signal only when we can image  both parent atoms. We accomplish that by sweeping the magnetic field back to the free atom side and dissociating the molecules. The temperature information therefore may not be accurately maintained during the diabatic sweeping process. With either dissociation method, we take \textit{in situ} absorption images to measure the number and spatial extent of the dimers as a function of hold time as shown in Fig.~\ref{fig:lifetime}. The width of the molecular cloud along the tightest confinement direction is about $5 \, \mathrm{\mu m}$, which is large compared to our imaging resolution of $1.5 \, \mathrm{\mu m}$. Both molecule width and imaging resolution are given in root-mean-square radius. When we image the atom component, we do a long time-of-flight to lower the optical density of the clouds.

%% figure S2

We measure the lifetime of the dimers with and without free atoms in the trap, denoted as $\tau_{\mathrm{tot}}$ and $\tau_{\mathrm{1b}}$ respectively. We fit the dimer population $N_{\mathrm{D}}$ with an exponential form and extract its $1/e$ lifetime as a function of scattering length $a$. We extract atom-dimer reaction coefficient from the rate equation:
\begin{align}
\label{eq:rateAD}
\frac{\mathrm{d}N_{\mathrm{D}}}{\mathrm{d}t} = & - \frac{N_{\mathrm{D}}}{\tau_{\mathrm{tot}}} = - \beta_{\mathrm{AD}} \langle n_\mathrm{A} \rangle N_{\mathrm{D}} - \frac{N_\mathrm{D}}{\tau_{\mathrm{1b}}},
\nonumber\\
& \beta_{\mathrm{AD}} = \frac{1}{\langle n_{\mathrm{A}} \rangle} \left( \tau_{\mathrm{tot}}^{-1} - \tau_{\mathrm{1b}}^{-1} \right),
\end{align}
where $\langle n_{\mathrm{A}} \rangle = N_{\mathrm{A}} \left[ m \bar{\omega}^{2} / \left( 4\pi k_{\mathrm{B}}T_{\mathrm{A}} \right) \right]^{3/2}$ is the mean density of atoms in a harmonic trap, where $N_{\mathrm{A}}$ is the atom number, $m$ is the atomic mass, $\bar{\omega}$ is the geometric mean trap frequency and $T_{\mathrm{A}}$ is the atom temperature. Dimers decay exponentially in time as long as atoms are the predominant population and collision processes between dimers are negligible. Note that $\beta_\mathrm{AD}$ here represents the result measured in a harmonic trap.

Throughout this work, we calibrate the magnetic field with rf spectroscopy. We use the most up-to-date Feshbach resonance calibration given by \cite{chapurin2019prl} for conversion from magnetic field to $a$. The uncertainty on scattering length amounts to about 5\% at around $20,000 \thinspace a_0$ and is negligible below $10,000 \thinspace a_0$.

\section{Finite Temperature Model for Atom-dimer Relaxation}
\label{sec:AD_model}

We follow the theoretical model introduced in \cite{helfrich2009epl} to incorporate finite temperature effects in the atom-dimer relaxation data. The decay rate of dimers is expressed as a thermal average of the rate coefficient at finite momentum,
\begin{align}
\label{eq:ensemble}
\frac{\mathrm{d}N_{\mathrm{D}}}{\mathrm{d}t} = & -\frac{1}{2\pi^3} \int_{0}^{\infty} r^2\mathrm{d}r \int_{0}^{\infty} P^2 \mathrm{d}P \int_{0}^{k_{\mathrm{D}}} k^2 \mathrm{d}k 
\nonumber\\ 
& \times \int_{-1}^{1} \mathrm{d}x \thinspace n_{\mathrm{A}}(p_{\mathrm{A}},r) n_{\mathrm{D}}(p_{\mathrm{D}},r) g(k),
\end{align}
where $\textbf{P} = \textbf{p}_{\mathrm{A}} + \textbf{p}_{\mathrm{D}}$ is total wave number of atom and dimer, $\textbf{k} = \frac{2}{3}\textbf{p}_{\mathrm{A}} - \frac{1}{3}\textbf{p}_{\mathrm{D}}$ is relative wave number of atom and dimer, and $x$ is the cosine of the angle between $\textbf{P}$ and $\textbf{k}$. We evaluate the integral of $k$ only up to the dimer break-up wave number $k_{\mathrm{D}} \approx 2/(\sqrt{3} a)$ for the model to be valid. We assume Boltzmann distribution of $n_{\mathrm{A}}$ and $n_{\mathrm{D}}$ since the temperature is always higher than $2 \times T_{\mathrm{c}}$ of atoms. 

The function $g(k)$ is related to the atom-dimer inelastic scattering cross section:
\begin{align}
\label{eq:crosssection}
g(k) & \equiv \frac{3 \hbar k}{2 m}  \sigma_{\mathrm{AD}}^{\mathrm{inelastic}}(k)
\nonumber\\ 
& = \frac{3 \hbar k}{2 m} \left( \sigma_{\mathrm{AD}}^{\mathrm{total}}(k) -  \sigma_{\mathrm{AD}}^{\mathrm{elastic}}(k) \right),
\end{align}
where $m$ is the atomic mass. We derive $\sigma_{\mathrm{AD}}^{\mathrm{total}}(k)$ and $\sigma_{\mathrm{AD}}^{\mathrm{elastic}}(k)$ from the atom-dimer scattering amplitude $f_{\mathrm{AD}}(k)$ that is expanded in series of the product $ka$ \cite{helfrich2009epl}. The short-range properties of atom-dimer scattering are encoded into $f_{\mathrm{AD}}(k)$ with $a_*$ and $\eta_*$ which are used as fitting parameters in our case. We then evaluate the integrals in Eqn.~\ref{eq:ensemble} and use the results to fit our $\beta_{\mathrm{AD}}$ data from $300 \, a_0$ to $2500 \, a_0$. The upper bound on $a$ guarantees that the temperature of the data always stays below one tenth of the dimer break-up energy. The lower bound is due to the assumption that this model will likely break down outside the universal regime. All the fitting parameters are summarized in Table.~\ref{table:S1}. There is a modest trend in $A_*$ with temperature, which we speculate is due to our imperfect modeling of the spatial overlap between molecular cloud and atomic cloud.

%% table S1
\begin{table*}
\begin{ruledtabular}
\begin{tabular}{ccccccc}
$T_{\mathrm{atom}}$ & $\langle n_{\mathrm{A}} \rangle$ & $\langle n_{\mathrm{D}} \rangle$ & $\bar{\omega}/2\pi$ & $a_{*}^{(1)}$ & $\eta_*$ & $A_*$ \\
\hline
nK & $10^{10} \thinspace \mathrm{cm^{-3}}$ & $10^{9} \thinspace \mathrm{cm^{-3}}$ & Hz & $a_{0}$ & & \\
\hline
$218(13)$ & $8.6(7)$ & $7.8$ & $42.7$ & $901(28)$ & $0.31(3)$ & $0.85(9)$ \\
$213(12)$ & $25.9(1.4)$ & $7.8$ & $42.7$ & $853(43)$ & $0.37(6)$ & $0.67(9)$ \\
$105(5)$ & $7.5(6)$ & $6.4$ & $24.4$ & $862(28)$ & $0.26(4)$ & $1.14(19)$ \\
$62(3)$ & $3.7(2)$ & $1.4$ & $17.1$ & $896(24)$ & $0.26(3)$ & $1.55(15)$
\end{tabular}
\end{ruledtabular}
\caption{Atom-dimer resonance measurement conditions and fit results. $T_\mathrm{atom}$ denotes the temperature of atoms. $\langle n_\mathrm{A} \rangle$ ($\langle n_\mathrm{D} \rangle$) is the density-weighted density of atoms (dimers). $\bar{\omega}$ is the geometric mean of trap frequency. The error bars on temperature and density represent the standard deviation of statistical noise.
}
\label{table:S1}
\end{table*}

\section{Density Dependence of Free-Atom Inelastic Collisions at Large Scattering Lengths}
\label{sec:AAA_density}

In the three-atom recombination measurements, absorption images are taken after a long time-of-flight to track both number loss and heating as a function of interrogation time. For certain range of $a$, we adapt the initial density of the cloud to keep the decay times within the convenient experimental range of tens of milliseconds to tens of seconds. The initial temperature of the cloud is constrained by this density for prohibiting quantum degeneracy. To describe the three-atom recombination decay process, we use the following rate equation for atom number $N$:
\begin{align}
\label{eq:threebodyrate}
\frac{1}{N} \frac{\mathrm{d}N}{\mathrm{d}t} = - \alpha_{\mathrm{bg}} - L_2 \langle n \rangle - L_3 \langle n^2 \rangle,
\end{align}
where $\langle n \rangle = N \left[ m \bar{\omega}^{2} / \left( 4\pi k_{\mathrm{B}}T \right) \right]^{3/2}$ is the spatially averaged density-weighted density, $\langle n^2 \rangle = N^2 \left[ m \bar{\omega}^{2} / \left( 2 \sqrt{3} \pi k_{\mathrm{B}}T \right) \right]^{3}$ is the averaged density-weighted square density, and $\alpha_{\mathrm{bg}}$ is the background loss rate which is calibrated for each individual condition at $60 \, a_0$. The temperature of the sample $T$ is being tracked together with atom number $N$ and then plugged into the rate equation in the form of cloud densities. The $d$-wave resonance at $63.6 \, \mathrm{G}$ is measured in a similar manner with the $L_3$ term being neglected in the corresponding small $a$ region.

%% figure S7
\begin{figure*}
\includegraphics[width=16cm,keepaspectratio]{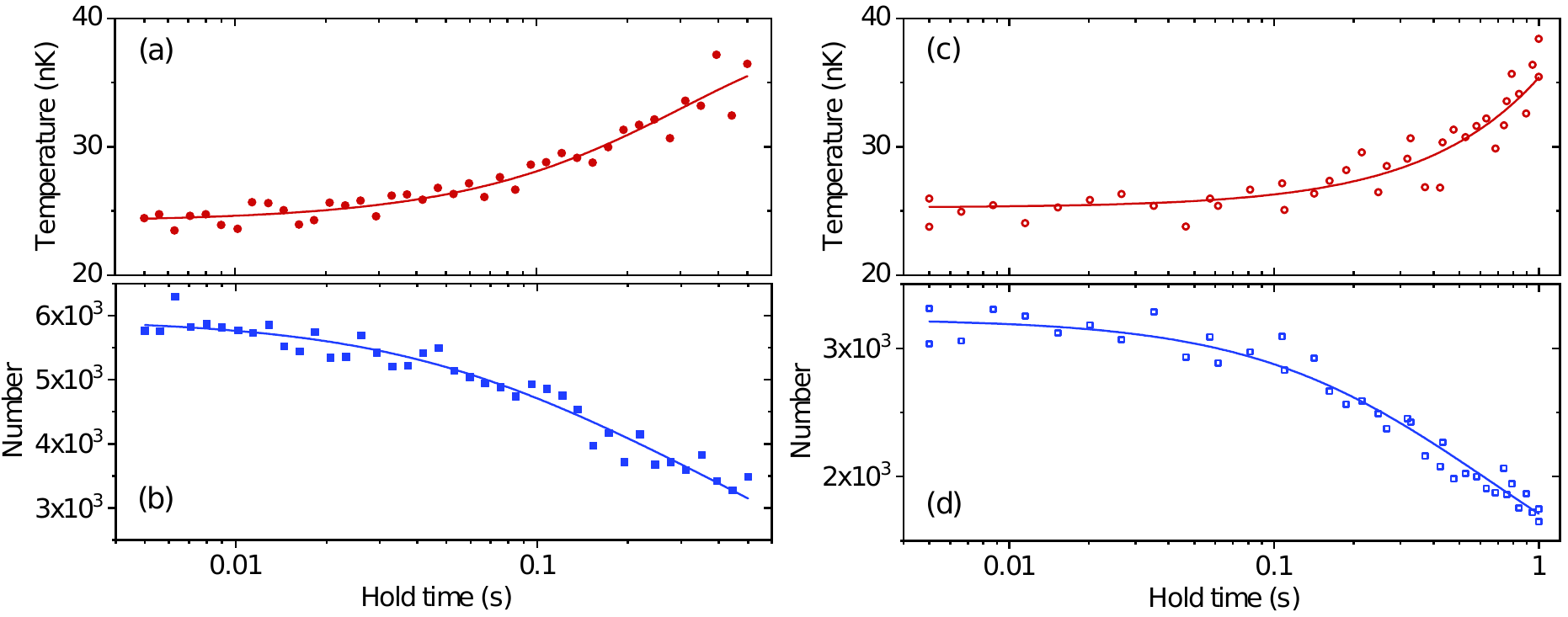}
\centering
\caption{Illustration of fitting protocol II that combines atom samples of two different initial densities but same time-averaged temperatures. (a)-(d) Time evolution of number (blue squares) and temperature (red circles) for the high density condition (solid symbols) and low density condition (open symbols) respectively. The data are taken at $3000 \, a_0$. Red solid lines are fitting curves for temperatures with polynomial. Blue solid lines are fitting curves for atom numbers with Eqn.~\ref{eq:threebodyrate}.
}
\label{fig:evolution} 
\end{figure*}

Ideally, we use relatively high density cloud while keeping the phase space density low so that the leading contribution to the loss is from the $L_3 \langle n^2 \rangle$ term. We fix $L_2$ to the theoretical values in our fitting function and float $L_3$ in this regime. We call it fitting protocol I. All data points above 210 nK in Fig.~5 of main text are analyzed by using fitting protocol I, and are consistent and readily interpretable up to $a \sim 2000 \, a_0$.

In large $a$ region, the contribution from two-body loss becomes notable and needs to be explored empirically. Although we can always extract both $L_2$ and $L_3$ by floating them in the rate equation, the fitting uncertainty is fairly large due to the inseparable time scales of two- and three-body losses for a single sample condition. To suppress the fitting error, we combine the number decay data that are taken on two samples of drastically different initial densities but identical temperatures and fit them simultaneously as depicted in Fig.~\ref{fig:evolution}. In this fitting protocol II, $L_2$ and $L_3$ are constrained to converge to the same respective values across data sets collected at varying initial densities, while atom numbers and temperatures from each data set are independent. 

%% figure S8
\begin{figure}
\includegraphics[width=8.4cm,keepaspectratio]{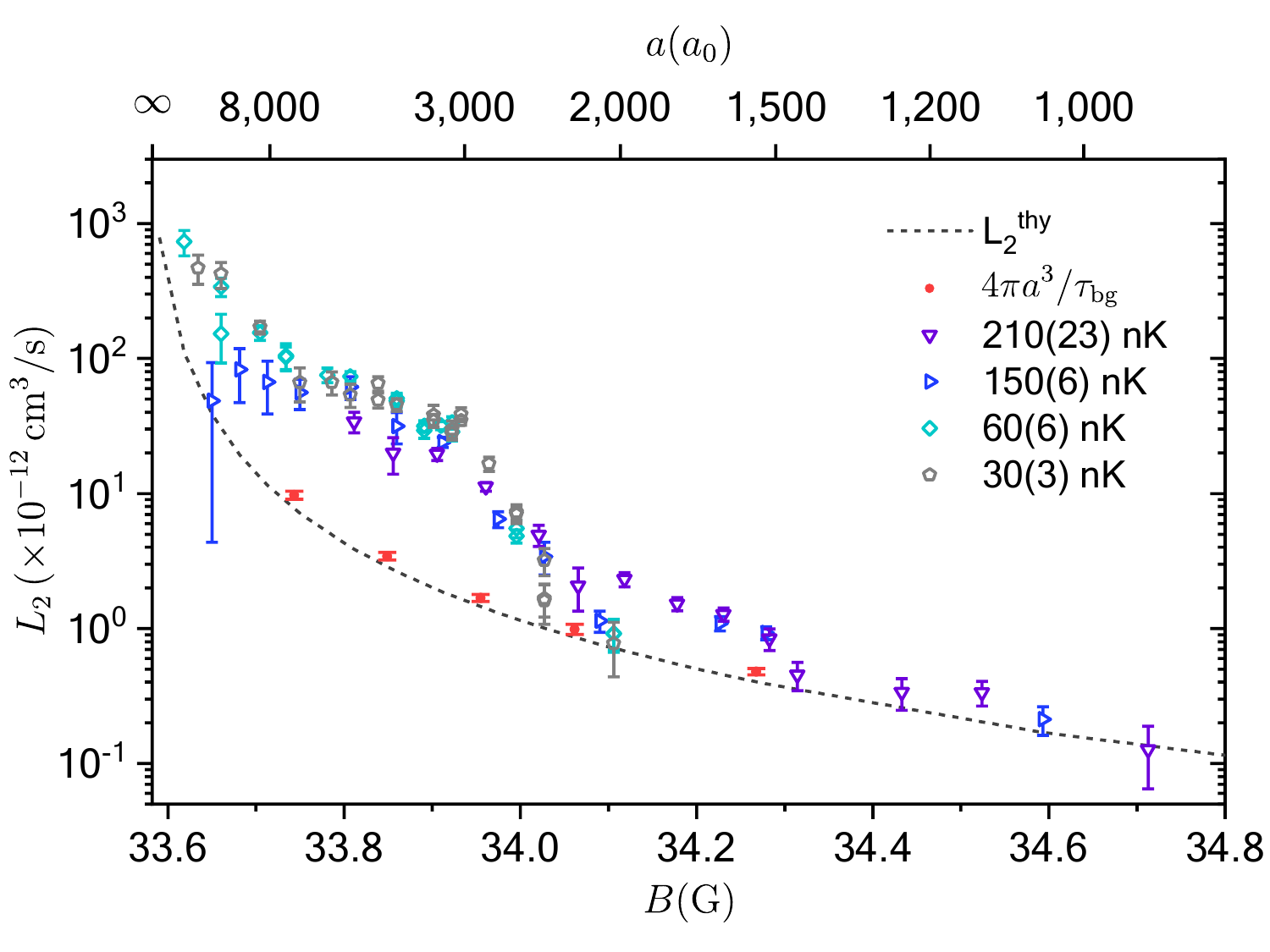}
\centering
\caption{Two-body inelastic scattering coefficient $L_2$ as function of $B$ or $a$. The open symbols are measurements taken with pure atom samples and extracted by using fitting protocol II. These experimental points presents a strong enhancement around $33.9 \, \mathrm{G}$ (or $3000 \, a_0$). This enhancement gradually diminishes for $a$ below $2000 \, a_0$. Solid red circles are deduced from shallow dimer lifetimes in the large $a$ limit by using the model from \cite{kohler2005prl}. The dashed line indicates the prediction from our refined two-body model on $sd$-wave basis.
}
\label{fig:2bpeak} 
\end{figure}

Above $2000 \, a_0$, we observe what presents as an enhanced $L_2$ behavior that can not be explained by the two-body coupled-channel theory as shown in  Fig.~\ref{fig:2bpeak}. This inconsistency is rather unexpected considering the perfect overlap between experiment and theory on the dimer intrinsic lifetime in the same region of $a$ (Fig.~\ref{fig:twobody}). Our detailed two-body model does not predict a resonance in the vicinity of $a = 3000 \thinspace a_0$. Our analysis protocol II yields not only the unexplained excess two-body decay at $a = 3000 \thinspace a_0$, but also scattered temperature dependence of the three-body recombination coefficient for $a > 1500 \thinspace a_0$. Taken together these results suggest that the overall explanation may come from some combination of uncharacterized systematic errors. We note that as $a$ becomes larger in magnitude, the binding energy of the dimer is no longer much greater than $k_\mathrm{B} T$, opening up other collisional channels. Also the energy released by three-body decay into the shallow dimer becomes no longer large enough to eject the decay products with absolute certainty \cite{langmack2013pra, rem2013prl}. Finally, the cloud begins to become collisionally opaque, giving rise to the possibility of decay products undergoing collisional cascade. A  sophisticated model that incorporates effects such as secondary collisions will be needed to track down the kinetics in this large positive $a$ regime.

\section{Finite Temperature Model for Three-body Recombination}
\label{sec:AAA_model}

We follow closely the S-matrix method introduced in \cite{braaten2008pra} to fit $L_3$ as a function of $a$ in order to account for the finite temperature effects. The three-body recombination coefficient is formulated as:
\begin{widetext}
\begin{align}
\label{eq:L3}
L_3 & \equiv N_{\mathrm{lost}} K_3(T) = N_{\mathrm{lost}} \frac{\displaystyle \int_{0}^{\infty} \mathrm{d}E \thinspace E^2 e^{-E/(k_{\mathrm{B}}T)} \left( K_{\mathrm{shallow}}(E) + K_{\mathrm{deep}}(E) \right)} {3! \displaystyle \int_{0}^{\infty} \mathrm{d}E \thinspace E^2 e^{-E/(k_{\mathrm{B}}T)}},
\end{align}
\end{widetext}
where $N_{\mathrm{lost}} = 3$ denotes the number of atoms being lost per collision event, the factor of $3!$ in the denominator accounts for the indistinguishable nature of three identical particles, and $E$ denotes the total energy of the three particles. $K_{\mathrm{shallow}}(E)$ and $K_{\mathrm{deep}}(E)$ represent the decay rate coefficients into the shallow dimer and deep dimers respectively. They can be expressed in terms of universal scaling functions $s_{12}(ka)$, $s_{22}(ka)$ and $s_{11}(ka)$: 
\begin{widetext}
\begin{align}
\label{eq:KofE}
K_{\mathrm{shallow}} & = \frac{144 \sqrt{3} \pi^2}{ (ka)^4 } \left( 1 - \bigg| s_{22} + 
\frac{s_{12}^2e^{2i\theta - 2\eta}}{1 - s_{11}e^{2i\theta - 2\eta} } \bigg|^2 - \frac{( 1 - e^{-4 \eta}) |s_{12}|^2 }{ | 1 - s_{11} e^{2i\theta - 2\eta} |^2}  \right) \frac{\hbar a^4}{m}, 
\nonumber\\ 
K_{\mathrm{deep}} & = \frac{144 \sqrt{3} \pi^2}{ (ka)^4 } \frac{( 1 - e^{-4 \eta}) \left( 1 - |s_{11}|^2 - |s_{12}|^2 \right) }{ | 1 - s_{11} e^{2i\theta - 2\eta} |^2 } \frac{\hbar a^4}{m},
\end{align}
\end{widetext}
where the momentum $k$ is related to $E$ through $E = \hbar^2 k^2/m$ with the atomic mass $m$, and the angle $\theta \equiv s_0 \thinspace \mathrm{ln}(a/a_+) = \pi/2 + s_0 \thinspace \mathrm{ln}(a/a_p)$ with $s_0 \approx 1.00624$ for three identical bosons. We include only the contribution of zero total orbital angular momentum in the above equations. The numeric values of $s_{ij}$ ($i,j = 1,2$) matrices are taken from \cite{braaten2008pra} in the intermediate $0 < ka < 1$ region. Prior to thermal averaging, $K_{\mathrm{shallow}}$ and $K_{\mathrm{deep}}$ are smoothly extrapolated to their threshold values in the limit of $ka \rightarrow 0$ where the numerical method loses its accuracy. The fit results are summarized in Table~\ref{table:S2}.

%% table S2
\begin{table*}
\begin{ruledtabular}
\begin{tabular}{cccccccc}
$T_{\mathrm{i}}$ & $\bar{T}$ & $\langle n \rangle_{\mathrm{i}}$ & $\bar{\omega}/2\pi$ & $a_{+}^{(0)}$ &  $a_{p}^{(0)}$ & $\eta_+$ or $\eta_p$ & $A_+$ \\
\hline
nK & nK & $10^{11} \thinspace \mathrm{cm^{-3}}$ & Hz & $a_{0}$ & $a_{0}$ & & \\
\hline
$370$ & $460(30)$ & $51.2(4.2)$ & $260.1$ & $246(6)$ & -- & $0.20(2)$ & $0.99(8)$ \\
$300$ & $410(40)$ & $26.4(2.8)$ & $79.4$ & -- & $893(28)$ & $0.20(0)$ & $0.74(2)$ \\
$150$ & $230(26)$ & $9.9(9)$ & $54.6$ & -- & $860(28)$ & $0.20(0)$ & $0.77(2)$
\end{tabular}
\end{ruledtabular}
\caption{Three-body recombination measurement conditions and fit results. $a_{+}^{(0)}$ and $a_{p}^{(0)}$ are obtained from the fits in a local range of the corresponding features so as to avoid presumptions about the Efimov period. The $\eta_p$ with fitting errors of zero are fixed parameters in the fit. $T_i$ and $\bar{T}$ denote initial temperature and temperature averaged over the interrogation time respectively. The error bar on $\bar{T}$ represents the variance during the interrogation time. $\langle n \rangle_i$ is the initial density-weighted density. The error bar on $\langle n \rangle_i$ represents the standard deviation of statistical noise. $\bar{\omega}$ is the geometric mean of trap frequency.}
\label{table:S2}
\end{table*}

%\bibliography{bib_file_supp}
